%% file: main.tex
\pdfoutput=1
\documentclass[12pt]{iopart}
\usepackage{graphicx}

\usepackage{amssymb}
\usepackage{amsthm}
\usepackage{bm}
\usepackage{graphicx}
\usepackage{ascmac}
\usepackage{color}
\usepackage{here}
\usepackage{iopams}
\usepackage[FIGTOPCAP]{subfigure}

\begin{document}

\def\mod{{\ \ \rm mod \,}}


\title[Eigenfunctions of Perron-Frobenius operator]{Eigenfunctions of the Perron-Frobenius operator and the finite-time Lyapunov exponents in uniformly hyperbolic area-preserving maps}
\author{Kensuke Yoshida$^{1,2}$, Hajime Yoshino$^1$, Akira Shudo$^1$ and Domenico Lippolis$^2$}
\address{$^1$ Department of Physics, Tokyo Metropolitan University, 
Minami-Osawa, Hachioji, Tokyo 192-0397, Japan}
\address{$^2$ Institute for Applied Systems Analysis, Jiangsu University, 212013 Zhenjiang, China}
\ead{domenico@ujs.edu.cn}

\begin{abstract}
	\input{0_abstract}
\end{abstract}
\submitto{\jpa}
\bibliographystyle{iopart-num}
\maketitle

\section{Introduction}\label{sec:introduction}
\input{1_introduction}

\section{The Perron-Frobenius operator}\label{sec:perronfrobenius}
\input{2_perronfrobenius}

\section{Numerical approximation of Perron-Frobenius operator I: the Ulam method}\label{sec:ulam}
\input{3_ulam}

\section{Numerical approximation of Perron-Frobenius operator II: Fokker-Planck operator method}\label{sec:fokkerplanck}
\input{4_fokkerplanck}

\section{Inhomogeneity of unstable manifolds}\label{sec:inhomgeneity_unst_mani}
\input{5_inhomogeneity_unst_mani}

\section{Finite-time Lyapunov exponent and second eigenfunction of Perron-Frobenius operator}\label{sec:finitetimelyapunov}
\input{6_finitetimelyapunov}

\section{Conclusion and Discussion}\label{sec:conclusion}
\input{7_conclusion}

\section*{Acknowledgements}
The authors are very grateful for Masato Tsujii, Fr\'ed\'eric Faure and Yuzuru Sato for their critical comments and helpful discussions. This work has been supported by JSPS KAKENHI Grants No. 15H03701 and No. 17K05583. DL is partially supported by the National Science Foundation of China, Young International Scientists (Grant No. 11750110416-1601190090).

\section*{References}
\bibliography{reference}
\end{document}

%% file: 0_abstract.tex
The subleading eigenvalues and associated eigenfunctions of the Perron-Frobenius operator for 2-dimensional area-preserving maps are numerically investigated. We closely examine the validity of the so-called Ulam method, a numerical scheme believed to provide eigenvalues and eigenfunctions of the Perron-Frobenius operator, both for linear and nonlinear maps on the torus. For the nonlinear case, the second-largest eigenvalues and the associated eigenfunctions of the Perron-Frobenius operator are investigated by calculating the Fokker-Planck operator with sufficiently small diffusivity. On the basis of numerical schemes thus established, we find that eigenfunctions for the subleading eigenvalues exhibit spatially inhomogeneous patterns, especially showing localization around the region where unstable manifolds are sparsely running. Finally, such spatial patterns of the eigenfunction are shown to be very close to the distribution of the maximal finite-time Lyapunov exponents. 

%% file: 1_introduction.tex
Uniformly hyperbolic systems, or, more loosely, strongly chaotic systems, exhibit unstable dynamics everywhere and the orbits asymptotically explore the whole phase space. Locally, or for short times, chaotic dynamical systems are characterized by local instability everywhere in the phase space, but, after a sufficiently long time, local information is averaged out, and a uniquely determined equilibrium distribution is achieved eventually.

This does not necessarily mean that no characteristic structures remain and everything proceeds homogeneously both in space and time, rather, there appear rich and varied structures worth examining even in a uniformly hyperbolic system. Recent decades have actually witnessed significant progress in characterizing spatial patterns and temporal behaviors of strongly chaotic systems. This includes developing a new concept such as the \textit{almost-invariant set} \cite{DJ1997, DJ1999}, which is proposed to capture regions in which the orbits tend to stay for a relatively long time compared to other regions. Another important finding as for the spatial structure hidden in the uniformly hyperbolic system is the position dependence of the escape rate \cite{KL2009, BY2011, Bun2012, APT2013}. The escape rate measures how fast the orbits initially launched in the phase space enter a hole punched at a certain place. For uniformly hyperbolic systems, the escape rate is supposed to be proportional to the hole size, and this is actually the case in the leading order. However, it was found that there is a substantial and actually observable correction originating from the periodic orbits of the system. Further analyses unveiled that the position dependence of the escape rate can be observed not only for mathematically well controlled situations \cite{KL2009, BY2011, Bun2012}, but also for more generic settings \cite{BD2007, APT2013, AB2017}.

The \textit{strange eigenmode} is an additional important topic, which should be mentioned in this context. In the field of the chaotic advection, it was reported that the distribution of the tracer particles stirred by external forces shows a fractal-like inhomogeneous pattern for a long time \cite{Pie1994}. Namely, the residence time of the tracer particles depends on the position in the state space, and the pattern is called the strange eigenmode \cite{Pie1994}. Note that the existence of the strange eigenmode for the advection-diffusion equation has been rigorously proved in the systems satisfying a suitable condition \cite{LH2004}. However, it is yet unclarified what geometric structures of the underlying deterministic system the strange eigenmode reflects \cite{Aetal2017}.

In this paper, we explore spatial and temporal properties of strongly chaotic systems, especially 2-dimensional area-preserving maps, by investigating the eigenvalues and eigenfunctions of the Perron-Frobenius operator. The Perron-Frobenius operator describes the time-evolution of distribution functions defined on the phase space. In particular, we focus on the leading sub-unit eigenvalue of the Perron-Frobenius operator and associated eigenfunction, which are referred to respectively as the second-largest eigenvalue and the second eigenfunction in our subsequent descriptions. As long as the second-largest eigenvalue is nonzero and is isolated from other eigenvalues, it controls the decay of correlations of observables, as well as the relaxation to the steady state \cite{Bal2000}. However, to the authors' knowledge, concrete examples providing reliable estimates of the second-largest eigenvalue of the Perron-Frobenius operator are limited \cite{Fro2007}, therefore it would be an important task to establish a numerical scheme capable of evaluating the isolated second-largest eigenvalues of any chaotic system.

In addition to the leading sub-unit eigenvalue, special attention will be paid to the nature of the second eigenfunction of the Perron-Frobenius operator. There is an indication for the relation between the second eigenfunction and almost-invariant sets \cite{DJ1999, Fro2008}, and the strange eigenmode also has a close link to them. Recalling that the Perron-Frobenius operator describes the time-evolution of distribution functions in the phase space, we can expect that spatial patterns of second eigenfunctions of the Perron-Frobenius operator carry significant information on spatial- or position dependence of the relaxation towards the equilibrium state.

We are also interested in the relation between the eigenfunctions of the Perron-Frobenius operator and those of the corresponding quantum time-evolution operator \cite{LSYY2020}, particularly with regard to localization patterns (scars) appearing in both.

The dynamical systems to be examined in this paper are well known hyperbolic systems. The first one is a linear torus automorphism $C:\mathbb{T}^2 \mapsto \mathbb{T}^2$, known as the Arnold cat map:
\begin{eqnarray}
  C:(x, y) \mapsto (x + y, x + 2y) \mod 1,
\end{eqnarray}
and another one is the so-called perturbed cat map given as
\begin{eqnarray}
  C_{\epsilon , \nu } :(x, y)&\mapsto C \circ F_{\epsilon, \nu}(x, y) \mod 1,
  \label{eq:def_pcm}
\end{eqnarray}
where
\begin{eqnarray}
  F_{\epsilon, \nu}: (x, y) &\mapsto \left( x - \frac{\epsilon}{\nu} \sin (2\nu \pi y), y \right).
\end{eqnarray}
Here the perturbation strength $\epsilon$ is assumed to be real and positive, and the perturbation frequency $\nu$ is a positive integer.

There are several works in the physics literature, where eigenvalues and eigenfunctions of the Perron-Frobenius operator have been calculated numerically \cite{BA2000, WHBMS2001, San2002, ES2010, ES2012, KW2016}. Apparently, numerical calculations can easily be implemented, for instance using the Ulam method introduced below. However, because of the highly pathological nature of eigenfunctions of the Perron-Frobenius operator in chaos, possibly coming from the fact that they are hyperfunctions \cite{BJ2002}, some straightforward numerical schemes can be unstable and thus unreliable. This is particularly so for the cat map $C$. Hence one of our main aims in this paper is to make clear to what extent one can access eigenfunctions of the Perron-Frobenius operator with numerical calculations. This point of view has not been seriously examined so far, although rigorous mathematical works strongly suggest that eigenfunctions, and eigenvalues as well, are not so easily handled. The answer depends on the functional spaces on which the Perron-Frobenius operator acts, and rigorous arguments cannot be developed without specifying such settings.

After establishing a reliable numerical scheme, we focus on the spatial signature of the second eigenfunctions. We pay particular attention to fractal-like- and at the same time inhomogeneously-distributed patterns observed in the perturbed cat map, and explore the features encoded in spatial patterns of the second eigenfunctions. The result clearly elucidates the position dependence of the relaxation rate, as recently discussed in the escape rate, almost invariant sets, and strange eigenmodes, etc.

The outline of the present paper is as follows. In \S \ref{sec:perronfrobenius}, we introduce the Perron-Frobenius operator. In \S \ref{sec:ulam} and \S \ref{sec:fokkerplanck}, we provide numerical results for second-largest eigenvalues and associated eigenfunctions of the Perron-Frobenius operator obtained first by using the Ulam method, and then by applying the Fokker-Planck to approximate the Perron-Frobenius operator. In \S \ref{sec:inhomgeneity_unst_mani}, we present numerical observations showing that eigenfunctions of the leading sub-unit eigenvalue exhibit spatially inhomogeneous patterns, and strong localization appears around the region where unstable manifolds are sparsely running. In \S \ref{sec:finitetimelyapunov}, we introduce the maximum finite-time Lyapunov exponent (mFT Lyapunov exponent, for short) to characterize the spatial pattern of the second eigenfunctions of the Perron-Frobenius operator. In \S \ref{sec:conclusion}, we draw conclusions and discuss our results. 

%% file: 2_perronfrobenius.tex
The Perron-Frobenius operator $\mathcal{L}$ for a map $f = C$ or $C_{\epsilon, \nu}$ is introduced as
\begin{eqnarray}
	\mathcal{L}\rho (\mathbf{x}) = \int_{\mathbb{T}^2}\delta (\mathbf{x} - f(\bar{\mathbf{x}})) \rho (\bar{\mathbf{x}}) \mathrm{d}\bar{\mathbf{x}},
	\label{eq:def_pf_1}
\end{eqnarray}
where $\mathbf{x} = (x,y) \in \mathbb{T}^2$ and the function $\rho (\mathbf{x})$ is defined on the phase space $\mathbb{T}^2$.

In the case of an area-preserving map, the Perron-Frobenius operator becomes unitary when it acts on the $L^2$ space \cite{Zei1995}. As a result, the Perron-Frobenius operator for an area-preserving map in $L^2$ does not have any non-zero sub-unit eigenvalues, when instead we are interested in the decay of correlations, as mentioned in the introduction. For that reason, we shall consider the Perron-Frobenius operator acting on other functional spaces. A typical setting in the mathematical treatment is to introduce a proper functional space for which the Perron-Frobenius operator is quasi-compact. A quasi-compact operator has a positive essential spectral radius, defined as the minimum radius of the disk such that all eigenvalues lying outside the disk are isolated and have at most finite degeneracy \cite{Bal2000}. In particular, the maximal eigenvalue is simple and equal to one, if the system is mixing, and its eigenfunction is a uniform distribution on $\mathbb{T}^2$, for an area-preserving map. The quasi-compactness is a necessary condition for the Perron-Frobenius operator to have a non-zero second-largest eigenvalue. It is not guaranteed that the desired functional space can be accessed within reach of numerical computations.

In the subsequent sections, we investigate the validity of numerical schemes approximating the Perron-Frobenius operator. We denote eigenvalues and the corresponding eigenfunctions of the Perron-Frobenius operator by $\sigma$ and $\psi$ respectively, and do not distinguish the ones obtained numerically from exact eigenvalues and eigenfunctions, if there is no ambiguity. In particular, if there exists an isolated second-largest eigenvalue, we denote it by $\sigma_2$ hereafter.

%% file: 3_ulam.tex
The Ulam method is one of the most commonly used numerical schemes for approximating the Perron-Frobenius operator \cite{Ula1960}. Let us consider a map $f$ and a partition of the phase space $\mathfrak{R} = \{ R_1, R_2, \dots \}$. In general, the partition $\mathfrak{R} = \{ R_i \}_{i=1,2,\dots, n}$ of the space $X$ is a family of subsets such that $X = \bigcup_{i=1}^{n} R_i$ and $R_i \cap R_j = \emptyset \; (i \neq j)$. In the Ulam method, one approximates the Perron-Frobenius operator for the map $f$ with the Ulam matrix, defined as
\begin{eqnarray}
	(L)_{ij} = \frac{m (R_i \cap f^{-1}(R_j))}{m (R_i)} \quad (i, j = 1, 2, \dots ),
	\label{eq:ulam_matrix}
\end{eqnarray}
where $f^{-1}$ is the inverse mapping of $f$ and $m(\cdot)$ the Lebesgue measure normalized in the phase space.

It was shown for some maps that the eigenfunction of $L$ associated with the eigenvalue 1 weakly converges to a measure which is absolutely continuous in the sense of Lebesgue or Sinai-Ruelle-Bowen, for an infinitely refined partition \cite{Li1976, BL1991, Fro1995, DZ1996, Bla1997, Fro1998}. The Ulam method has also been used to approximate sub-unit eigenvalues and their eigenfunctions \cite{ES2010, ES2012, KW2016}. On the other hand, rigorous results on the Ulam method are rather limited \cite{BSTV1997, BK1998, Fro2007}, and the results for hyperbolic linear maps are briefly mentioned in our subsequent arguments.

In the following, we will apply the Ulam method to the cat map and the perturbed cat map in order especially to explore to what extent the Ulam method works in calculating sub-unit eigenvalues and the corresponding eigenfunctions for the Perron-Frobenius operator.

We numerically construct the Ulam matrix in the following way. First put $N_j$ initial points randomly chosen in the $j$-th partition $R_j \in \mathfrak{R}$ where $N_j$ is taken to be proportional to the area of $R_j$, and then count the points whose backward iterations are contained in the $i$-th partition $R_i$. We denote the number of such points by $N_{ij}$:
\begin{eqnarray}
	N_{ij} = \# \{ \mathbf{x}^{(n)} \in R_j \; (n=1,2,\dots,N_j) \vert f^{-1}(\mathbf{x}^{(n)}) \in R_i\}.
\label{eq:count_points}
\end{eqnarray}
The $(i,j)$-component of the Ulam matrix is set to $ \left(L\right)_{ij}  = \displaystyle \frac{N_{ij}}{\sum_i N_{ij}}$.

\subsection{Cat map case}
\subsubsection{Markov partition case}
In this subsection, we apply the Ulam method for the unperturbed cat map $C$ by using the Markov partitions to generate the Ulam matrix. Given a partition $\mathfrak{R} = \{ R_1, R_2, \dots , R_N \}$, where each region $R_i$ has a boundary $\partial R_i = \partial ^S R_i \bigcup \partial ^U R_i$, and $\partial ^S R_i$ and $\partial ^U R_i$ respectively in the stable and unstable manifolds, $\mathfrak{R}$  is a Markov patition if $C(\bigcup_{i=1}^N \partial ^S R_i) \subset \bigcup_{i=1}^N \partial ^S R_i$ and $C^{-1}(\bigcup_{i=1}^N \partial ^U R_i) \subset \bigcup_{i=1}^N \partial ^U R_i$ \cite{Bowen75}. In particular, we will consider the two different types of Markov partitions, which are respectively referred to as the Adler-Weiss (AW, for short) and Brini-Siboni-Turchetti-Vaienti (BSTV, for short) partition \cite{AW1967, BSTV1997}. We denote the AW partition by $\mathfrak{M}_{A}^0$ and the BSTV partition by $\mathfrak{M}_{B}^0$.

\begin{figure}[H]
	\begin{minipage}{\hsize}
		\begin{center}
			\includegraphics[width = \hsize, bb = -100 0 900 300]{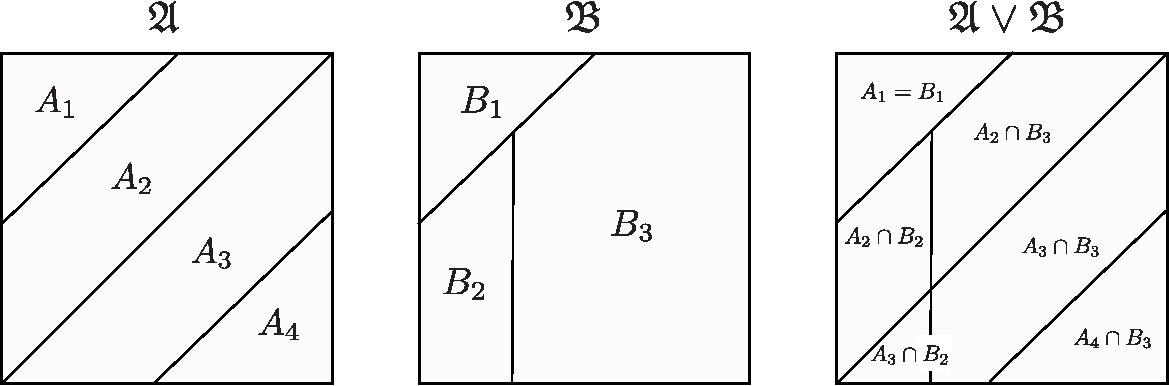}
		\end{center}
	\end{minipage}
	\caption{
		Illustration of the partition $\mathfrak{A}\vee \mathfrak{B}$, which is	generated by the partitions $\mathfrak{A}$ and $\mathfrak{B}$.
	}
	\label{fig:vee_product}
\end{figure}

\begin{figure}[H]
	\subfigure[\hspace{2cm}]{
		\begin{minipage}{0.32\hsize}
			\begin{center}
				\includegraphics[width = \hsize, bb = 0 0 350 345.6]{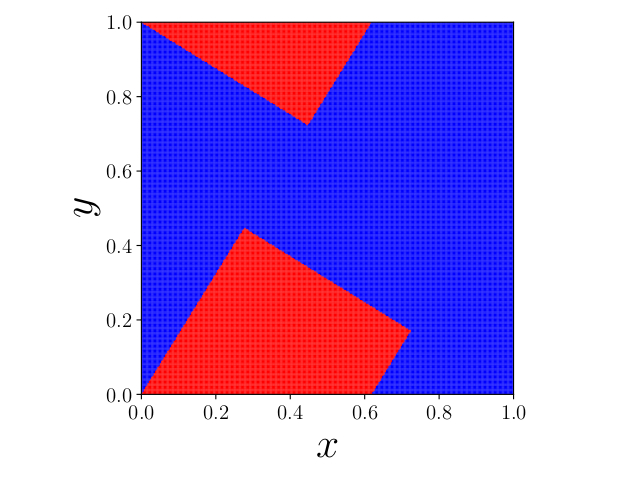}
			\end{center}
		\end{minipage}
	}
	\subfigure[\hspace{2cm}]{
		\begin{minipage}{0.32\hsize}
			\begin{center}
				\includegraphics[width = \hsize, bb = 0 0 350 345.6]{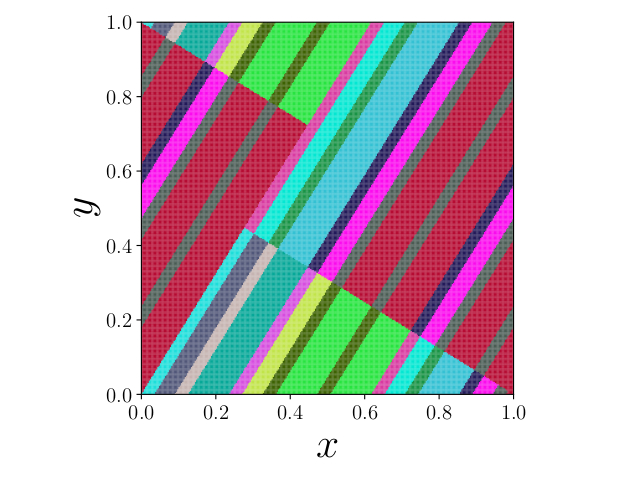}
			\end{center}
		\end{minipage}
	}
	\subfigure[\hspace{2cm}]{
		\begin{minipage}{0.32\hsize}
			\begin{center}
				\includegraphics[width = \hsize, bb = 0 0 350 345.6]{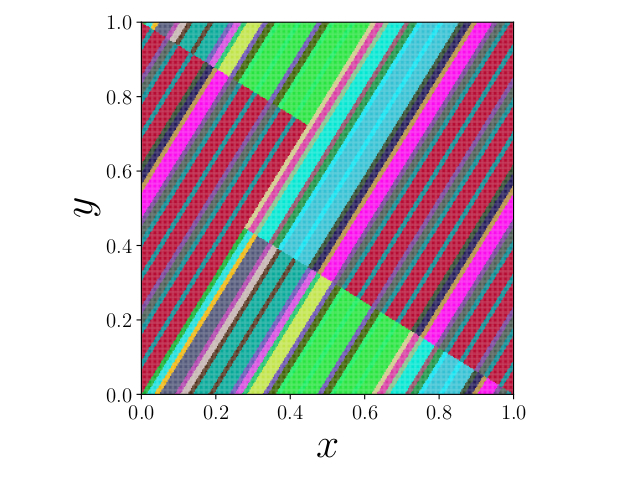}
			\end{center}
		\end{minipage}
	}
	\subfigure[\hspace{2cm}]{
		\begin{minipage}{0.32\hsize}
			\begin{center}
				\includegraphics[width = \hsize, bb = 0 0 350 345.6]{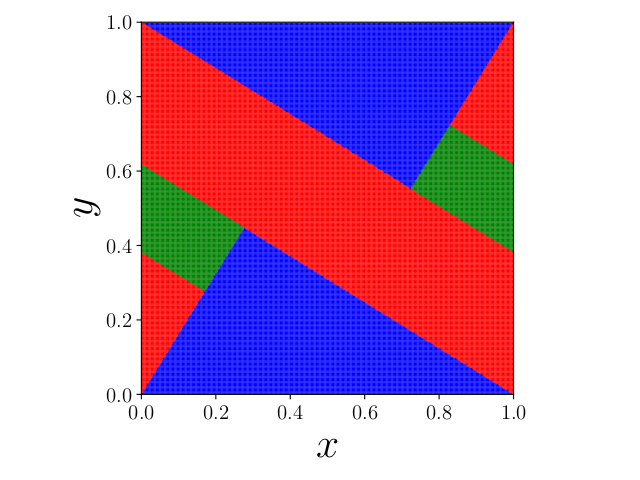}
			\end{center}
		\end{minipage}
	}
	\subfigure[\hspace{2cm}]{
		\begin{minipage}{0.32\hsize}
			\begin{center}
				\includegraphics[width = \hsize, bb = 0 0 350 345.6]{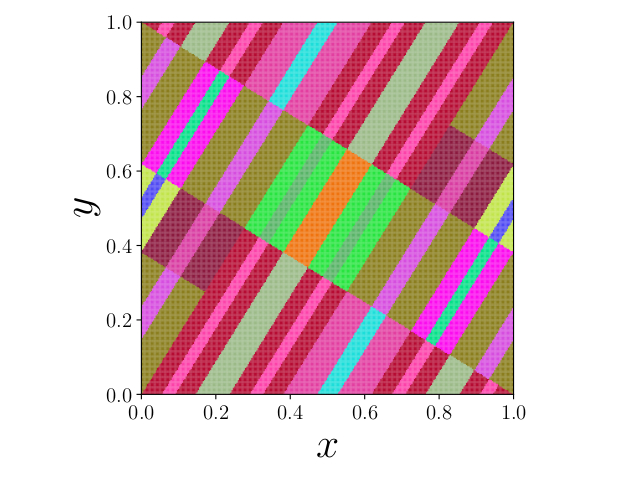}
			\end{center}
		\end{minipage}
	}
	\subfigure[\hspace{2cm}]{
		\begin{minipage}{0.32\hsize}
			\begin{center}
				\includegraphics[width = \hsize, bb = 0 0 350 345.6]{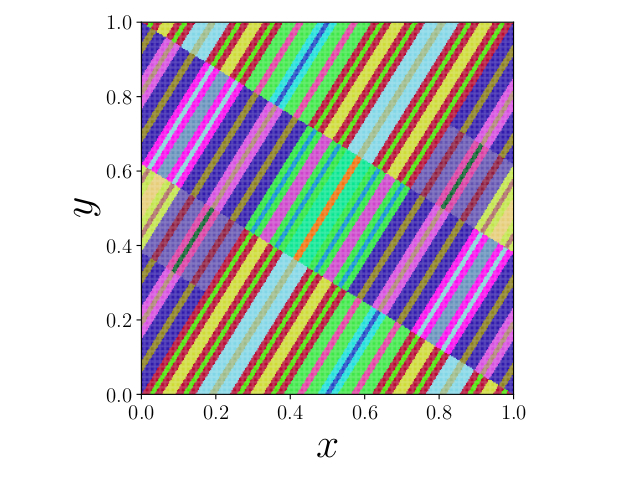}
			\end{center}
		\end{minipage}
	}
	\caption{
		(a)-(c) The AW partition $\mathfrak{M}^{t_{\mathrm{max}}}_A$ with (a) $t_{\mathrm{max}} = 0$, (b) $t_{\mathrm{max}} = 3$ and (c) $t_{\mathrm{max}} = 4$. (d)-(f) The BSTV partition $\mathfrak{M}^{t_{\mathrm{max}}}_B$ with (d) $t_{\mathrm{max}} = 0$, (e) $t_{\mathrm{max}} = 3$ and (f) $t_{\mathrm{max}} = 4$.
	}
	\label{fig:initial_markov_partition}
\end{figure}

For a Markov partition $\mathfrak{M}$,  either $\mathfrak{M}_A$ or $\mathfrak{M}_B$, the finer partition obtained by
\begin{eqnarray}
	\label{eq:finer_markov_partition}
	\mathfrak{M} \vee C(\mathfrak{M}) = \{ E_{\mathfrak{M}}(\mathbf{x}) \cap E_{C(\mathfrak{M})}(\mathbf{x}) ; \mathbf{x} \in \mathbb{T}^2 \} ,
\end{eqnarray}
is also a Markov partition where $C$ is the cat map. Here $E_{\mathfrak{M}}(\mathbf{x})$ denotes the elements of the partition $\mathfrak{M}$ containing the point $\mathbf{x}$. An example illustrating the procedure given in equation (\ref{eq:finer_markov_partition}) is shown in figure \ref{fig:vee_product}. The partition $\mathfrak{A}\vee \mathfrak{B}$ represents the finest partition generated from the partitions $\mathfrak{A}=\{ A_1, A_2, A_3, A_4 \}$ and $\mathfrak{B}=\{ B_1, B_2, B_3 \}$.
The finer partition generated iteratively up to the $t_{\mathrm{max}}$-th image of the partition $C^{t_{\mathrm{max}}}(\mathfrak{M})$ is denoted by
\begin{eqnarray}
	\mathfrak{M}^{t_{\mathrm{max}}} = \mathfrak{M} \vee C(\mathfrak{M}) \vee C^2(\mathfrak{M}) \dots \vee C^{t_{\mathrm{max}}}(\mathfrak{M}).
\end{eqnarray}
Note that $\mathfrak{M}^{t_{\mathrm{max}}}$ is also a Markov partition. We call $\mathfrak{M}_{A({\rm resp.} B)}^{t_{\mathrm{max}}}$ the $t_{\mathrm{max}}$-th AW(resp. BSTV) partition. As shown in figure \ref{fig:initial_markov_partition}, the AW partition $\mathfrak{M}_A^{0}$ is composed of two rectangles, and does not have any spatial symmetry whereas the BSTV partition $\mathfrak{M}_B^{0}$ has three rectangles, and is point symmetric in the torus $\mathbb{T}^2$.

\begin{figure}[H]
	\subfigure[\hspace{4cm}]{
		\begin{minipage}{0.5\hsize}
			\includegraphics[width = \hsize, bb = 0 0 460.8 345.6]{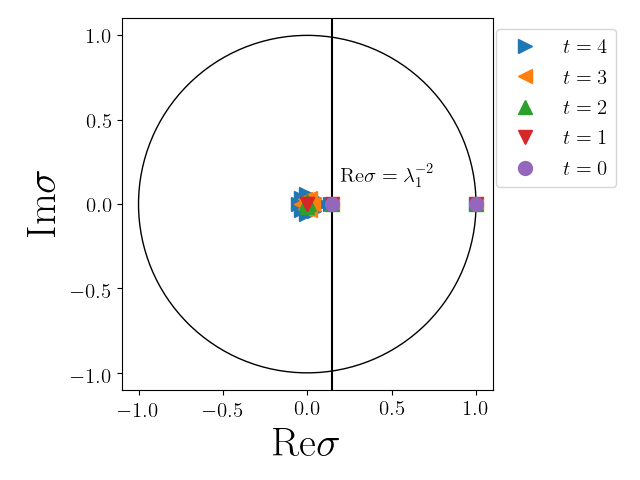}
		\end{minipage}
	}
	\subfigure[\hspace{4cm}]{
		\begin{minipage}{0.5\hsize}
			\includegraphics[width = \hsize, bb = 0 0 460.8 345.6]{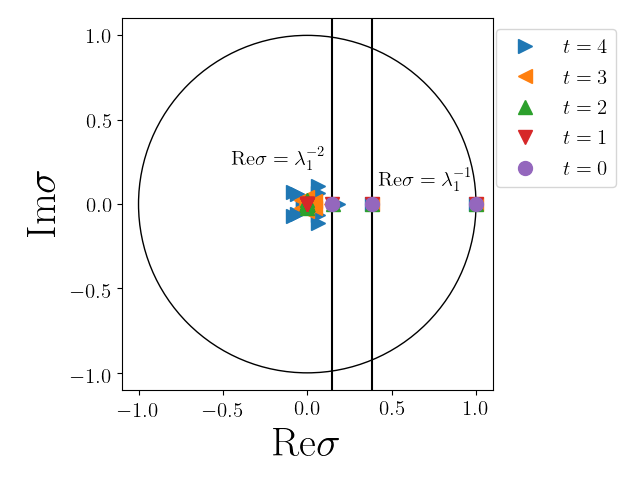}
		\end{minipage}
	}
	\caption{
		Eigenvalues of the Ulam matrix generated with the Markov partitions (a) AW partition and (b) BSTV partition.
	}
	\label{fig:ulam_markov_eigenval}
\end{figure}

\begin{figure}[H]
	\subfigure[\hspace{4cm}]{
		\begin{minipage}{0.5\hsize}
			\includegraphics[width = \hsize, bb = 0 0 460.8 345.6]{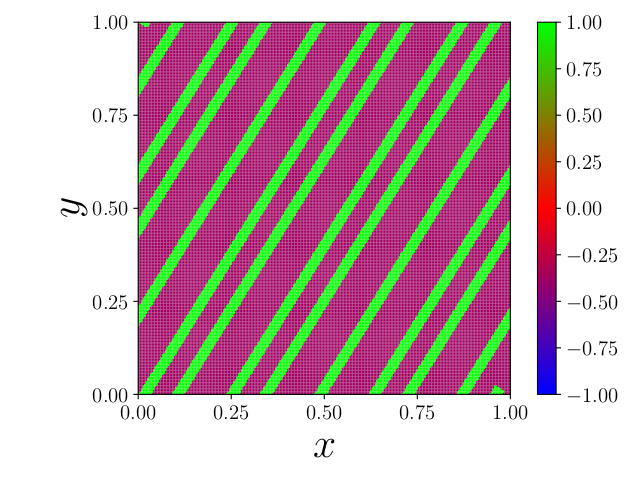}
		\end{minipage}
	}
	\subfigure[\hspace{4cm}]{
		\begin{minipage}{0.5\hsize}
			\includegraphics[width = \hsize, bb = 0 0 460.8 345.6]{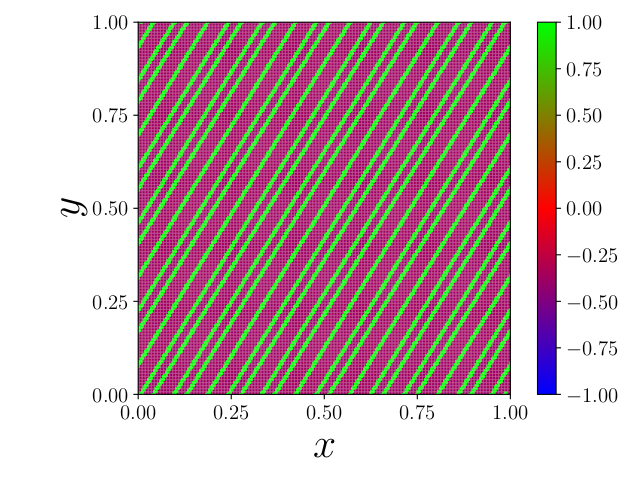}
		\end{minipage}
	}
	\subfigure[\hspace{4cm}]{
		\begin{minipage}{0.5\hsize}
			\includegraphics[width = \hsize, bb = 0 0 460.8 345.6]{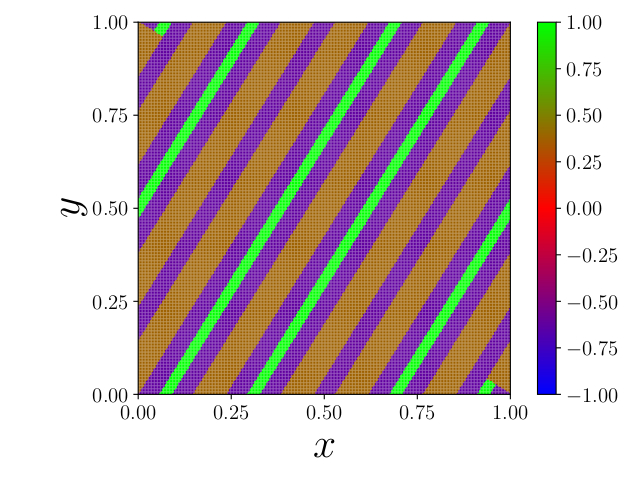}
		\end{minipage}
	}
	\subfigure[\hspace{4cm}]{
		\begin{minipage}{0.5\hsize}
			\includegraphics[width = \hsize, bb = 0 0 460.8 345.6]{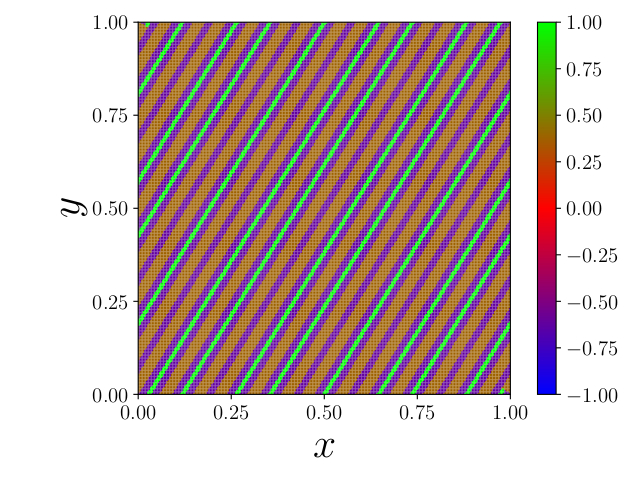}
		\end{minipage}
	}
	\subfigure[\hspace{4cm}]{
		\begin{minipage}{0.5\hsize}
			\includegraphics[width = \hsize, bb = 0 0 460.8 345.6]{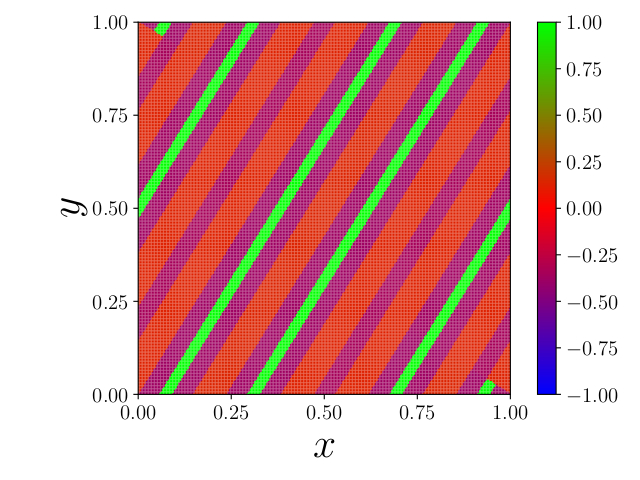}
		\end{minipage}
	}
	\subfigure[\hspace{4cm}]{
		\begin{minipage}{0.5\hsize}
			\includegraphics[width = \hsize, bb = 0 0 460.8 345.6]{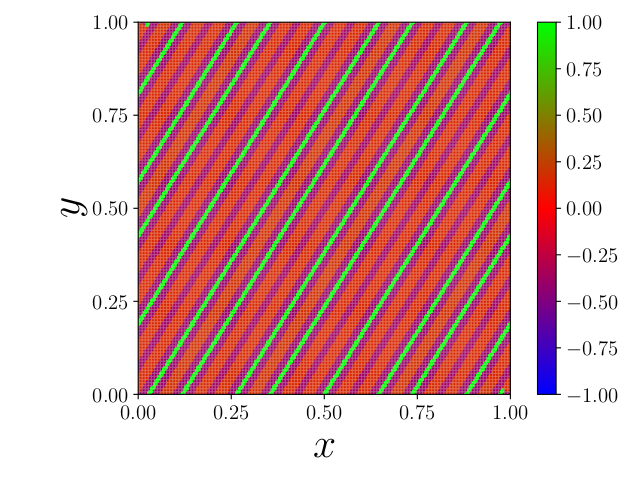}
		\end{minipage}
	}
	\caption{
		(a)-(b) The second eigenfunction generated using the AW partition with (a) $t_{\mathrm{max}} = 3$ and (b) $t_{\mathrm{max}} = 4$.	(c)-(d) The second eigenfunction generated using the BSTV partition with (c) $t_{\mathrm{max}} = 3$ and (d) $t_{\mathrm{max}} = 4$. (e)-(f) The third eigenfunction generated using the BSTV partition with (e) $t_{\mathrm{max}} = 3$ and (f) $t_{\mathrm{max}} = 4$.
	}\label{fig:ulam_cat_markov_second_eigenvec}
\end{figure}

Let  $\mathcal{L}_{A({\rm resp.} B)}$ be the Ulam matrix generated with the AW(resp. BSTV) partition. Brini et al. proved, by extending the results in \cite{AW1967}, that the sets of the eigenvalues of $\mathcal{L}_A$ and $\mathcal{L}_B$ are $\{ 0, (\lambda^{(C)})^{-2}, 1\}$ and $\{ 0, (\lambda^{(C)})^{-2}, (\lambda^{(C)})^{-1}, 1\}$, respectively \cite{BSTV1997}. The numerically obtained eigenvalues of $\mathcal{L}_A(\mathrm{resp.} B)$ are shown in figure \ref{fig:ulam_markov_eigenval}. One  can find that the second-largest eigenvalues are $(\lambda^{(C)})^{-2}(\mathrm{resp.} (\lambda^{(C)})^{-1})$, independent of the cardinality of the partitions. The complex eigenvalues placed around the origin $(0,0)$ are spurious, that appear as a result of numerical artifacts. All spurious eigenvalues approach the origin of the complex plane with the increase of the initial points $N_j$ [cf. equation (\ref{eq:count_points})].
Here $\lambda^{(C)} = (3 + \sqrt{5})/2$ is the larger eigenvalue of the Jacobi matrix $\partial C_i/\partial x_j$. We call $\lambda^{(C)}$ the stability multiplier of the cat map.	The multiplicity of the non-zero eigenvalues of $\mathcal{L}_{A({\rm resp.} B)}$ is 1 \cite{BSTV1997}. To our knowledge, it has not been clarified what functional space the Perron-Frobenius operator approximated by the Ulam matrix acts on, when the cardinality of the partition is taken to be infinity \cite{BKL2002}. On the other hand, the Perron-Frobenius operator for the area-preserving map becomes unitary if it is defined as the operator acting on the Hilbert space $L^2(\mathbb{T}^2)$, and in such a case the second-largest eigenvalue turns out to be $0$~\cite{Zei1995}. Even if one restricts the functional space on which the Perron-Frobenius operator acts to smooth functions, sub-leading eigenvalues, which are called the {\it Ruelle-Pollicott resonances}, are shown to be all zero \cite{San2002,FR2006}. These results imply that the resulting eigenvalues depend on the method we use and the functional space we consider.
The second eigenfunctions of $\mathcal{L}_A$ or $\mathcal{L}_B$ obtained by performing numerical calculations are presented in figure \ref{fig:ulam_cat_markov_second_eigenvec}. The second eigenfunction of $\mathcal{L}_A$ appears to be a binary valued function, and the second eigenfunction of $\mathcal{L}_B$ seems a ternary valued function within the scale of the color bar of the figure, strongly suggesting that they are truly binary or ternary, although no rigorous proofs exist, to our knowledge. We show the eigenfunction associated with the third-largest eigenvalue of $\mathcal{L}_B$ in figure \ref{fig:ulam_cat_markov_second_eigenvec}(e)-(f). Although the third-largest eigenvalue of $\mathcal{L}_B$ is identical to the second-largest eigenvalue of $\mathcal{L}_A$, the eigenfunction associated with the third-largest eigenvalue of $\mathcal{L}_B$ looks a ternary valued function within the scale of the color bar of the figure. Therefore, even if the eigenvalues take the same value, the eigenfunctions of the Ulam matrices can exhibit different patterns depending on the choice of the Markov partition.

On the other hand, one can find some properties common to eigenfunctions obtained from the AW partition and BSTV partition. The pattern of the second eigenfunction becomes finer with increase in $t_{\mathrm{max}}$ (compare the left and the right panel in figures \ref{fig:ulam_cat_markov_second_eigenvec}(a) and (b)). The second eigenfunction of the Ulam matrix generated with $\mathfrak{M}_{A({\rm resp.} B)}^{t_{\mathrm{max}}}$ shows the pattern with a smaller scale of that obtained by $\mathfrak{M}_{A(B)}^{t_{\mathrm{max}}-1}$. Because of this property, we can expect that the second eigenfunction tends to a fractal function as $t_{\mathrm{max}} \rightarrow \infty$.

The second eigenfunction is smooth along unstable manifolds and wildly oscillating in the direction of stable manifolds. Such a pattern is commonly observed in numerically obtained eigenfunctions associated with sub-unit eigenvalues of the Ulam matrix for the standard map \cite{Fro2007}. Furthermore, such patterns can be obtained not only by the Ulam method but also by other methods \cite{BA2000, WHBMS2001}. In particular, it was reported that in the quadbaker map, that is also uniformly hyperbolic, piecewise linear, and area-preserving, the absolute value of the second eigenfunction generated by a certain partition smoothly changes along the unstable manifold, and it wildly oscillates along the stable manifold. It was rigorously proved that it converges to the second eigenfunction of the Perron-Frobenius operator on a suitable functional space as the partition tends to an infinite refinement \cite{Fro2007}. Note that in \cite{Fro2007} the author used a partition consisting of equally-sized squares, whose sides are parallel to the stable and unstable manifolds.

\subsubsection{Regular partition case}
We here investigate how the calculation based on the Ulam method with a uniform discretization of the phase space works for the unperturbed cat map $C$. The equally-spaced partition will hereafter be referred to as the {\it uniform partition}. Previously, there has been an attempt in the same direction \cite{ES2012}. As will be illustrated below, however, the Ulam method with uniform partitions is not stable enough to obtain eigenvalues and the corresponding eigenfunctions, and thus we have to say that it does not necessarily lead to conclusive results.

We here construct the uniform partition $\mathfrak{R}_M$ by dividing the phase space into the $M \times M$ equally spaced squared boxes and prepare the Ulam matrix associated with the partition $\mathfrak{R}_M$. Note that for the unperturbed cat map $C$ and any $M$, each matrix element takes the value of either $0$ or $1/4$ \cite{BK1998}, which makes the calculation of matrix elements free from numerical errors.

\begin{figure}[H]
	\subfigure[\hspace{4.5cm}]{
		\begin{minipage}{0.5\hsize}
			\includegraphics[width = \hsize, bb = 0 0 460.8 345.6]{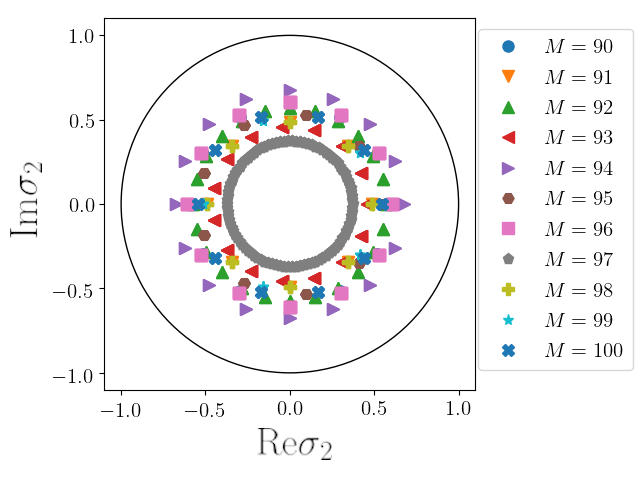}
		\end{minipage}
	}
	\subfigure[\hspace{5cm}]{
		\begin{minipage}{0.5\hsize}
			\includegraphics[width = \hsize, bb = 0 0 460.8 345.6]{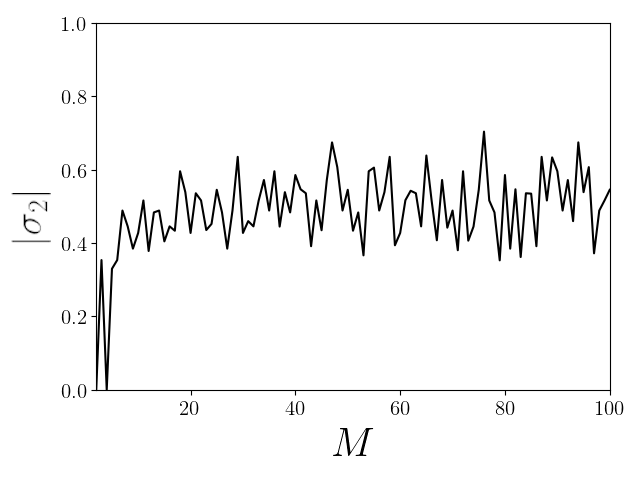}
		\end{minipage}
	}
	\caption{(a) Second-largest eigenvalues of the Ulam matrix generated with the uniform partition. The outer solid circle is the unit circle.
		For $M=97$, numerous second-largest eigenvalues are overlapped and form a circle.
		(b) The absolute value of the second-largest eigenvalue on the real axis generated with the uniform partition plotted as a function of the partition size $M$.}
	\label{fig:ulam_cat_eigenval}
\end{figure}

In figure \ref{fig:ulam_cat_eigenval}(a), we compute the second-largest eigenvalues for $\mathcal{L}_M$. Notice that there appear several second-largest eigenvalues for every given $M$, taking the same absolute value within numerical errors, and the number of second-largest eigenvalues depends on $M$. Note that for $M = 97$, numerous second-largest eigenvalues appear so as to form a circle in the complex plane. Additionally, as seen in figure \ref{fig:ulam_cat_eigenval}(b), the absolute value of the second-largest eigenvalue largely fluctuates even with increase in $M$, and does not seem to converge to an asymptotic value, up to $M=100$. Such a fluctuating behavior of the eigenvalues has been already reported in \cite{BK1998}.

It was proved that the Ulam matrix generated with the uniform partition has a self-similar structure, meaning that $\mathcal{L}_{kM}$ is composed of the combination of $\mathcal{L}_M$ where $k$ is a positive integer \cite{BK1998}. Although $\mathcal{L}_M$ has such a clear structure, it is only proved that all of the eigenvalues of the $\mathcal{L}_{M}$ are contained in the set of the eigenvalues of the $\mathcal{L}_{kM}$ for any given positive integer $k$ \cite{BK1998}. We cannot conclude whether the second-largest eigenvalue of $\mathcal{L}_{M}$ converges to a certain value in the limit of $M \rightarrow \infty$ or not.

\begin{figure}[H]
	\subfigure[\hspace{4cm}]{
		\begin{minipage}{0.5\hsize}
			\includegraphics[width = \hsize, bb = 0 0 460.8 345.6]{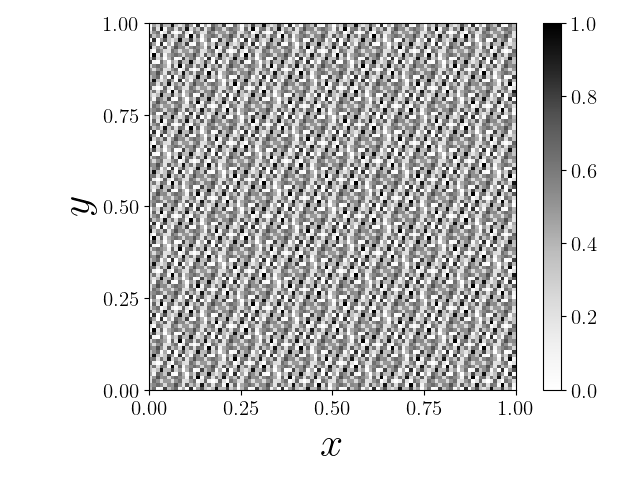}
		\end{minipage}
	}
	\subfigure[\hspace{4cm}]{
		\begin{minipage}{0.5\hsize}
			\includegraphics[width = \hsize, bb = 0 0 460.8 345.6]{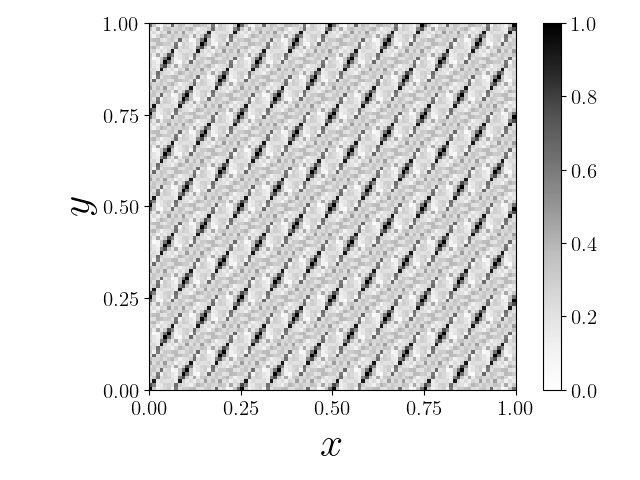}
		\end{minipage}
	}
	\caption{
		Second eigenfunctions (absolute value) for the cat map obtained
		using the Ulam matrix generated with the uniform partition. (a) and (b) are associated with different second-largest eigenvalues. $M = 100$ was taken.}
	\label{fig:ulam_cat_second_eigenvec}
\end{figure}

\begin{figure}[H]
	\begin{center}
		\subfigure[\hspace{10cm}]{
			\begin{minipage}{0.7\hsize}
				\begin{center}
				\includegraphics[width=\hsize ]{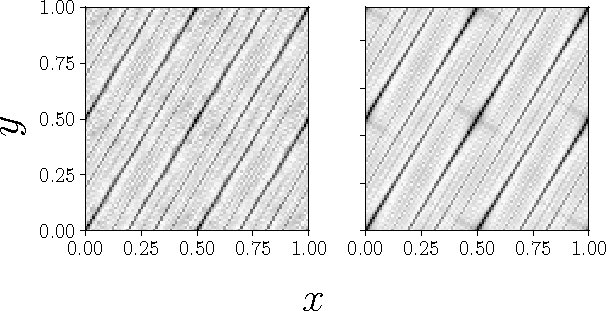}
				\end{center}
			\end{minipage}
		}
	\subfigure[\hspace{10cm}]{
			\begin{minipage}{0.7\hsize}
				\includegraphics[width=\hsize]{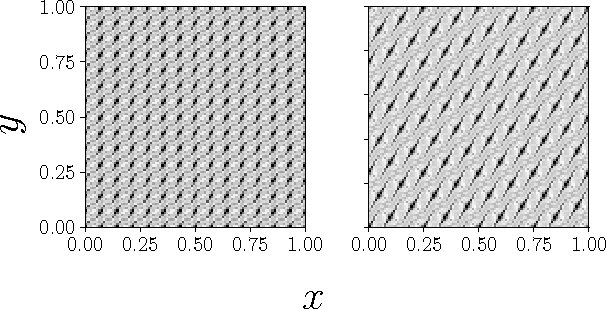}
			\end{minipage}
		}
		\subfigure[\hspace{10cm}]{
			\begin{minipage}{0.7\hsize}
				\includegraphics[width=\hsize]{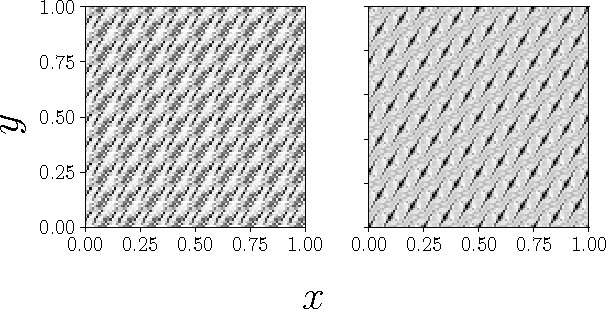}
			\end{minipage}
		}
	\end{center}
	\caption{Pairs of second eigenfunctions (absolute values) whose
		differences are relatively small.  (a) $(M, M') = (92, 94)$ with $\Delta \psi_{i,j}^{(M,M')}\simeq 0.17208$, (b) $(M, M') = (98, 100)$ with $\Delta \psi_{i,j}^{(M,M')}\simeq 0.44639$ and (c) $(M, M') = (99, 100)$ with $\Delta \psi_{i,j}^{(M,M')}\simeq 0.46836$.
	}
	\label{fig:ulam_cat_second_eigenvec_pairs_M100}
\end{figure}

Since the entries of the Ulam matrix generated with the uniform partition are all real but the matrix is non-symmetric, the eigenfunctions become complex-valued, in general. We hereafter plot their magnitudes. In figure \ref{fig:ulam_cat_second_eigenvec}, we first present a pair of second eigenfunctions, whose eigenvalues have the same absolute value but different phases. It was proved that eigenfunctions for $\mathcal{L}_M$ show repeating patterns \cite{BK1998}, however it is not clear whether the mathematical statement presented in \cite{BK1998} justifies periodic patterns observed in figure~\ref{fig:ulam_cat_second_eigenvec}, and also the results reported in \cite{ES2012}.
Since, as seen above, the numerical evaluation of the eigenvalues is rather unstable, it would be reasonable to expect that the same is true for eigenfunctions. In order to verify this, we introduce a norm evaluating the difference of the eigenfunctions obtained using different grid numbers $M$ and $M'$ as
\begin{eqnarray}
	\Delta \psi_{i,j}^{(M, M')} : = \frac{1}{\bar{M}\times \bar{M}}\sum_{k = 1}^{\bar{M}\times \bar{M}}\vert \psi_{i}^{(M)}(k) - \psi_{j}^{(M')}(k) \vert .
\end{eqnarray}
Here $\psi_{i}^{(M)}(k)$ denotes the value of a second eigenfunction supported on a finer uniform grid of $\bar{M}\times \bar{M}$ squares at the $k$-th interval $(1 \leq k \leq \bar{M} \times \bar{M})$. The index $i$ runs in the set of the second eigenfunctions, and $\bar{M}$ stands for the least common multiple of $M$ and $M'$. Obviously, $\Delta \psi_{i,j}^{(M, M')}$ is non-negative and equal to 0 if $M = M'$. If $\min_{i,j} \Delta \psi_{i,j}^{(M, M')}$ is close to zero, then $\mathcal{L}_{M}$ and $\mathcal{L}_{M'}$ have second eigenfunctions with a similar spatial pattern.
\begin{figure}[H]
	\subfigure[\hspace{4cm}]{
		\begin{minipage}{0.5\hsize}
			\includegraphics[width = \hsize, bb = 0 0 460 400]{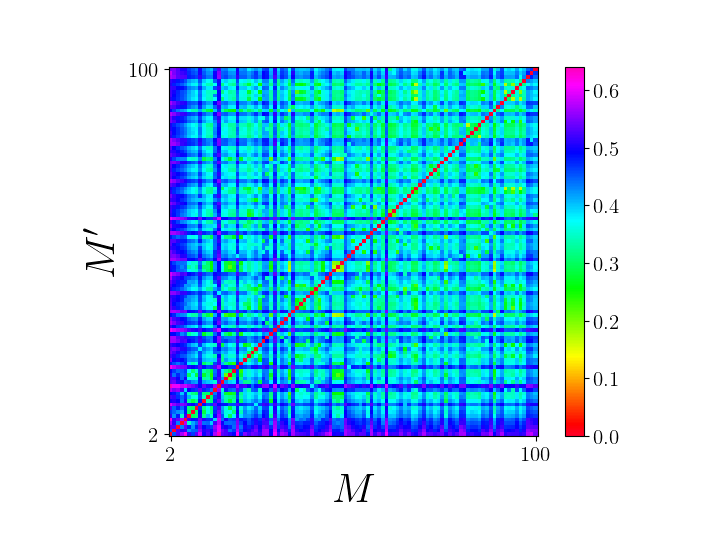}
		\end{minipage}
	}
	\subfigure[\hspace{4cm}]{
		\begin{minipage}{0.5\hsize}
			\includegraphics[width = \hsize, bb = 0 0 460.8 400.6]{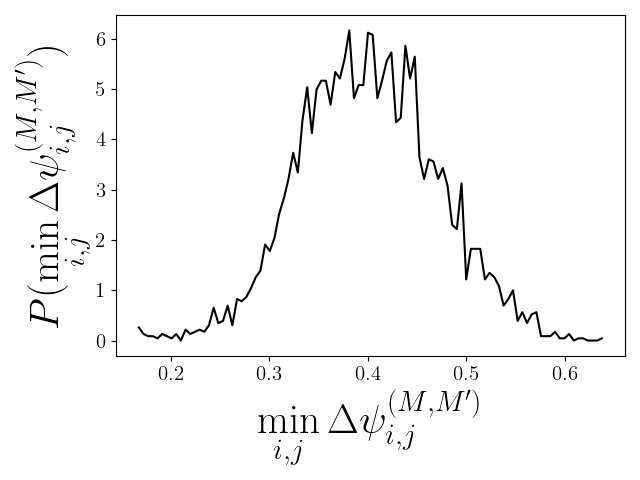}
		\end{minipage}
	}
		\caption{
			(a) The plot of $\displaystyle \min_{i,j}\Delta \psi_{i,j}^{(M, M')}$. (b) The normalized distribution of $\displaystyle \min_{i,j}\Delta \psi_{i,j}^{(M, M')}$ for $M > M'$.
		}
		\label{fig:ulam_cat_second_eigenvec_diff}
\end{figure}
The result of figure \ref{fig:ulam_cat_second_eigenvec_diff}(a) tells us that the values of $\Delta \psi_{i,j}^{(M, M')}$ are non-negligibly small,	except on the trivial line ($M = M'$). As shown in figure \ref{fig:ulam_cat_second_eigenvec_diff}(b), the distribution of $\min_{i,j}\Delta \psi_{i,j}^{(M, M')}$ exhibits a bell-shaped profile, and its mean value is around 0.4. In figure \ref{fig:ulam_cat_second_eigenvec_pairs_M100}(a), we present a pair of second eigenfunctions, whose difference $\Delta \psi_{i,j}^{(M,M')}$ is relatively small compared with the mean value of $\min_{i,j}\Delta \psi_{i,j}^{(M, M')}$. These spatial patterns happen to be similar to each other, but such a pair is rarely found. As shown in figure \ref{fig:ulam_cat_second_eigenvec_pairs_M100}(b) and (c), the typical pairs of second eigenfunctions, whose differences $\Delta \psi_{i,j}^{(M,M')}$ are close to the mean value of $\min_{i,j}\Delta \psi_{i,j}^{(M, M')}$, show totally different spatial patterns. From these calculations, we must conclude the Ulam method with uniform partitions is not suitable to calculate both eigenvalues and eigenfunctions in the case of the unperturbed cat map $C$.
\subsection{Perturbed cat map case}
Next we apply the Ulam method to the perturbed cat map $C_{\epsilon, \nu}$. It was shown that the perturbed cat map is topologically conjugate to the cat map if the condition $0 < \epsilon < (\sqrt{5} - 1) /4\sqrt{2}\pi \approx 0.069$ is satisfied \cite{Arn1985, DD1995, DC2003}. Furthermore, we show that the eigenvalues of the Jacobian for the perturbed cat map are both real and positive for $0 \leq \epsilon < 1/2\pi \approx 0.159$ (see \S \ref{sec:inhomgeneity_unst_mani}). Hereafter, we fix $\nu = 1$ but the following numerical results are essentially the same in the $\nu > 1$ cases.
It is not easy or at most numerically achievable to construct the Markov partition for the perturbed cat map. We here use the uniform partition $\mathfrak{R}_M$ introduced in the previous subsection.

\begin{figure}[H]
	\subfigure[\hspace{4cm}]{
		\begin{minipage}{0.5\hsize}
			\includegraphics[width = \hsize, bb = 0 0 460.8 345.6]{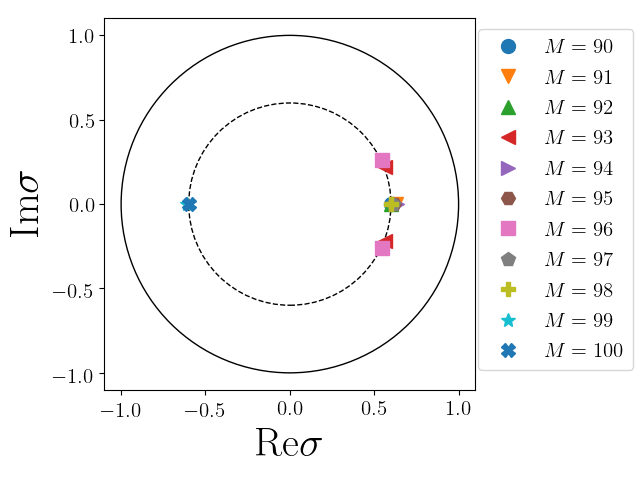}
		\end{minipage}
	}
	\subfigure[\hspace{4cm}]{
		\begin{minipage}{0.5\hsize}
			\includegraphics[width = \hsize, bb = 0 0 460.8 345.6]{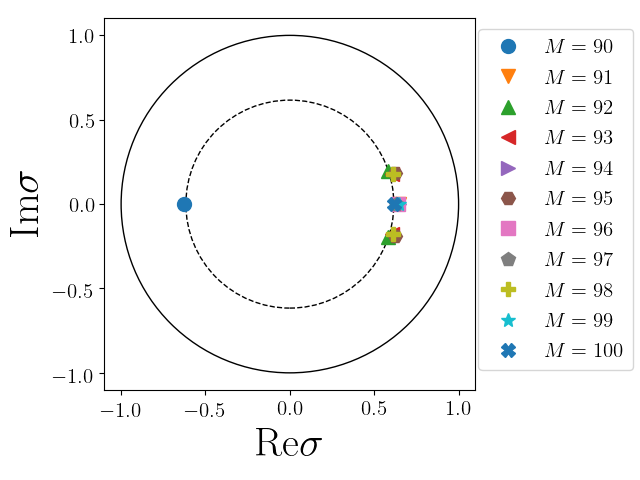}
		\end{minipage}
	}
	\subfigure[\hspace{4cm}]{
		\begin{minipage}{0.5\hsize}
			\includegraphics[width = \hsize, bb = 0 0 460.8 345.6]{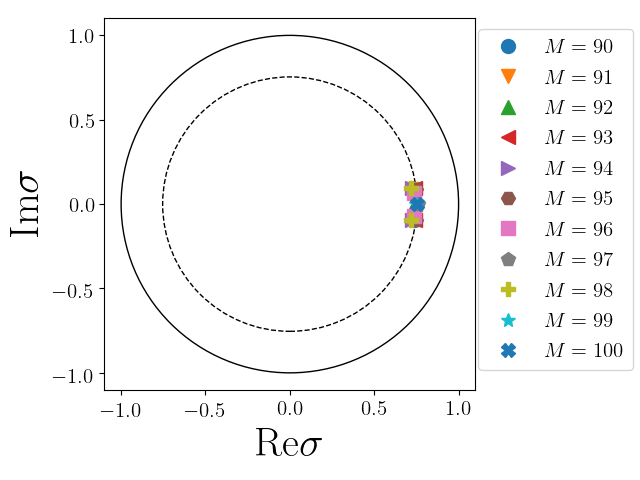}
		\end{minipage}
	}
	\subfigure[\hspace{4cm}]{
		\begin{minipage}{0.5\hsize}
			\includegraphics[width = \hsize, bb = 0 0 460.8 345.6]{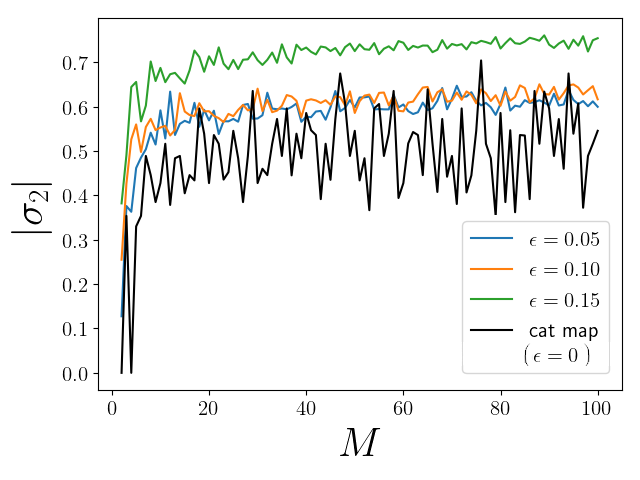}
		\end{minipage}
	}
	\caption{
		Second-largest eigenvalues of the Ulam matrices for the perturbed cat map with (a) $\epsilon = 0.05$, (b) $\epsilon = 0.1$ and (c) $\epsilon = 0.15$. $\nu = 1$ is taken. The broken line is the circle whose radius equals $\vert \sigma_2 \vert$ for $M = 100$. The outer solid curve represents the unit circle. (d) The absolute value of a second-largest eigenvalue of the Ulam matrix as a function of $M$. $\nu = 1$ is taken.
	}	\label{fig:ulam_perturbed_cat_eigenval}
\end{figure}

Figures~\ref{fig:ulam_perturbed_cat_eigenval}(a)-(c) display the second-largest eigenvalues of the Ulam matrix. In each case, there appear  multiple second-largest eigenvalues having the same modulus but different phases. As the perturbation strength $\epsilon$ increases we find a tendency of the multiple second-largest eigenvalues to come close to each other.

\begin{figure}[H]
	\subfigure[\hspace{4cm}]{
		\begin{minipage}{0.5\hsize}
			\includegraphics[width = \hsize, bb = 0 0 460.8 345.6]{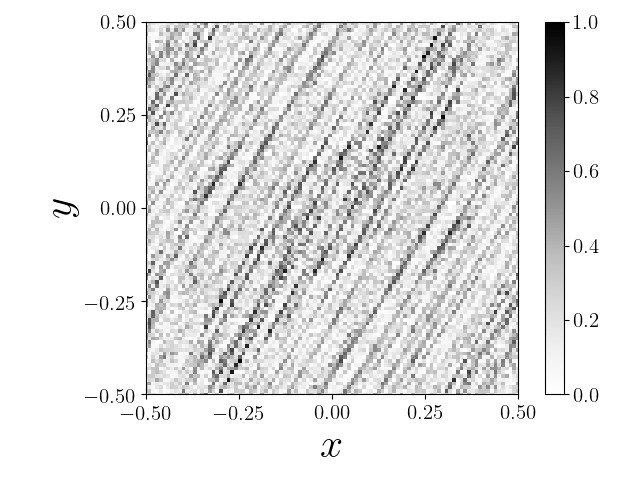}
		\end{minipage}
	}
	\subfigure[\hspace{4cm}]{
		\begin{minipage}{0.5\hsize}
			\includegraphics[width = \hsize, bb = 0 0 460.8 345.6]{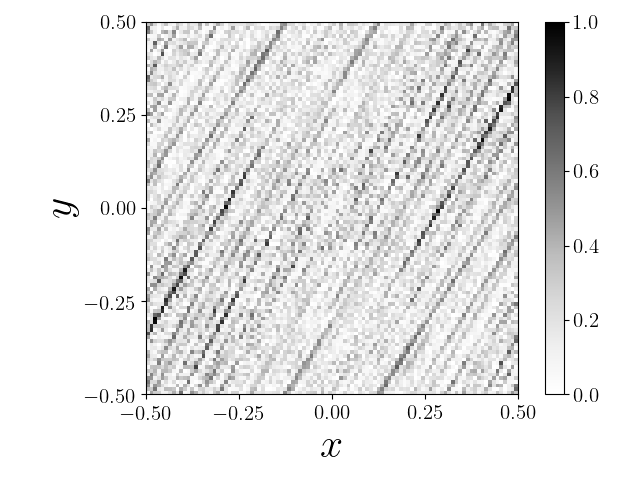}
		\end{minipage}
	}
	\subfigure[\hspace{4cm}]{
		\begin{minipage}{0.5\hsize}
			\includegraphics[width = \hsize, bb = 0 0 460.8 345.6]{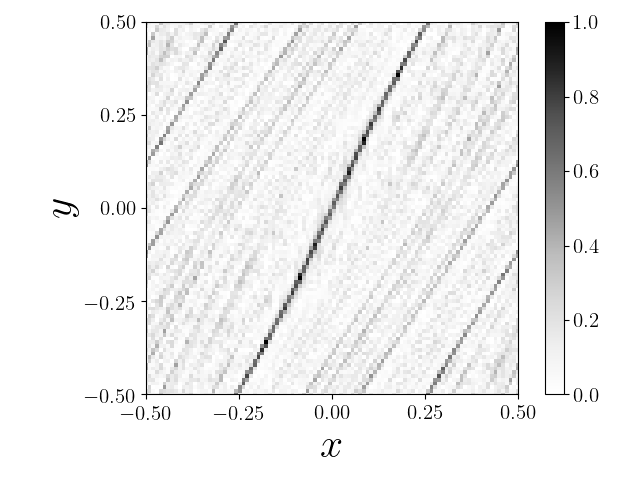}
		\end{minipage}
	}
	\subfigure[\hspace{4cm}]{
		\begin{minipage}{0.5\hsize}
			\includegraphics[width = \hsize, bb = 0 0 460.8 345.6]{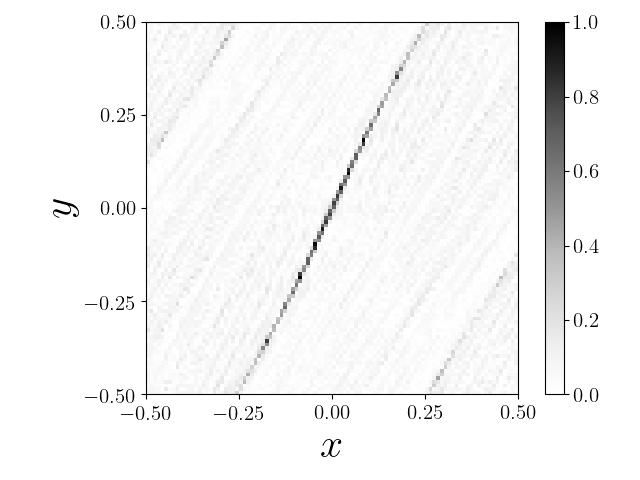}
		\end{minipage}
	}
	\subfigure[\hspace{4cm}]{
		\begin{minipage}{0.5\hsize}
			\includegraphics[width = \hsize, bb = 0 0 460.8 345.6]{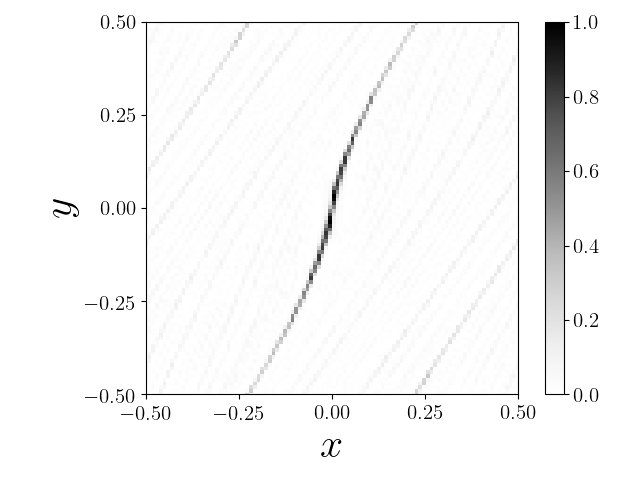}
		\end{minipage}
	}
	\subfigure[\hspace{4cm}]{
		\begin{minipage}{0.5\hsize}
			\includegraphics[width = \hsize, bb = 0 0 460.8 345.6]{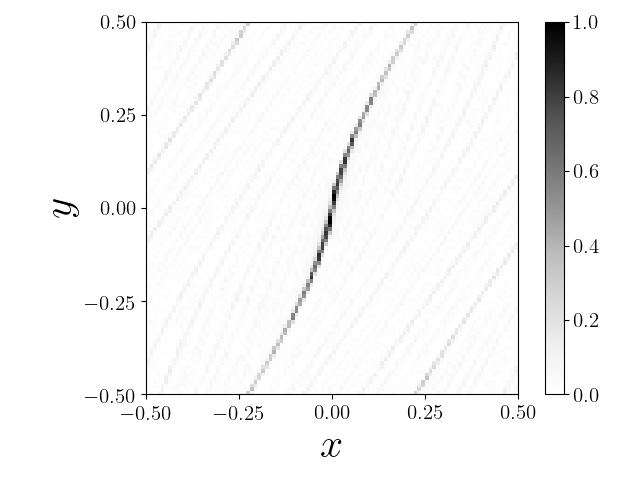}
		\end{minipage}
	}
	\caption{
		Magnitudes of the	second eigenfunctions of the Ulam matrices for the perturbed cat map with (a) $\epsilon = 0.05$, $M = 99$, (b) $\epsilon = 0.05$, $M = 100$, (c) $\epsilon = 0.1$, $M = 99$, (d) $\epsilon = 0.1$, $M = 100$, (e) $\epsilon = 0.15$, $M=99$, and (f) $\epsilon = 0.15$, $M=100$.  $\nu = 1$ is taken all through.
	}	\label{fig:ulam_perturbed_cat_second_eigenvec}
\end{figure}

As shown in figure \ref{fig:ulam_perturbed_cat_eigenval}(d), the convergence of the second-largest eigenvalues as a function of the size $M^2 \times M^2$ of the Ulam matrix improves compared to the case of the unperturbed cat map $C$, although fluctuations are still noticeable. A similar scenario is found in the patterns of the corresponding eigenfunctions. As seen in figure \ref{fig:ulam_perturbed_cat_second_eigenvec}, spatial patterns depend on $M$. We notice, however, that the difference of spatial patterns between the case with $M=99$ and $M=100$ is even smaller as the perturbation strength $\epsilon$ is increased. In addition, it should be noted that localized regions become more clearly recognizable, and computational stability improves as $\epsilon$ increases. This localized pattern reminds us of the scarring phenomenon in the quantum chaotic system \cite{Hel1984}. As will be closely discussed in the next section, localized regions appear along unstable manifolds and so this will become an important signature characterizing the pattern of second eigenfunctions.

In closing this section, we briefly summarize the results obtained so far. The Ulam method is one of the most common numerical schemes to compute the eigenvalues and eigenfunctions of the Perron-Frobenuis operator for chaotic maps. Our analyses have revealed that the result depends sensitively on the choice of the grid employed to construct the Ulam matrix in the case of the pure cat map, although it is considered to be a simplest Anosov map and expected to play the role of a prototype model for strongly chaotic systems. We should thus conclude that the eigenvalues and eigenfunctions of the Perron-Frobenuis operator for the unperturbed cat map are delicate objects in the sense of numerical calculations. As for the perturbed cat map, on the other hand, we have tested the Ulam method with uniform partitions, and found that it shows relatively reasonable convergence, compared to the case of the unperturbed cat map, although fluctuations of the second-largest eigenvalue still remain even when the size of the Ulam matrix is significantly increased.

%% file: 4_fokkerplanck.tex
The Ulam method is unstable for calculating eigenvalues and eigenfunctions of the Perron-Frobenius operator for the unperturbed cat map $C$, whereas it works reasonably well for the perturbed cat map $C_{\epsilon,\nu}$. In order to confirm that the characteristic localization patterns observed in the second eigenfunctions shown in figure~\ref{fig:ulam_perturbed_cat_second_eigenvec} are inherent of the perturbed cat map, we approximate the evolution operator using a Fourier basis. One reason for this choice is, as shown below, that it yields better convergent results, and can be used to examine the Perron-Frobenius operator by taking the noiseless limit of the Fokker-Planck operator. Another reason is that it is worth investigating physically in its own right since it naturally introduces a cutoff of spatial resolution, which is inevitable when the system is subject to external noise or coupled with a heat bath, etc. Introducing noise, we replace the Perron-Frobenius operator with the Fokker-Planck operator. 

The Fokker-Planck operator for the map $f = C$ or $C_{\epsilon, \nu}$ is defined as follows \cite{BA2000, TC2003, FR2006}.
\begin{eqnarray}
	\mathcal{L}_{\Delta}\rho (\mathbf{x}) = \int_{\mathbb{T}^2}\sum_{\mathbf{k} \in \mathbb{Z}^2}e^{2\pi \mathrm{i}\mathbf{k} \cdot (f(\mathbf{x}) - \bar{\mathbf{x}}) - \mathbf{k}^2\Delta} \rho(\bar{\mathbf{x}})\mathrm{d}\bar{\mathbf{x}},
	\label{eq:fokker_planck_operator}
\end{eqnarray}
where $\mathbf{k}$ denotes the wavenumber vector and $\Delta$ a positive real parameter controlling the variance of the Gaussian integral kernel of the operator, which represents the amplitude of the noise, and makes the eigenvalues and eigenfunctions converge faster by suppressing the high-frequency Fourier modes. We call $\Delta$ diffusivity hereafter.

If one restricts the functions on which the Fokker-Planck operator $\mathcal{L}_{\Delta}$ acts to the Hilbert space $L^2(\mathbb{T}^2)$, the Fokker-Planck operator $\mathcal{L}_{\Delta}$ becomes a matrix of entries equal to the transition probabilities between Fourier modes $\mathbf{k}$ and $\mathbf{q}$ \cite{TC2003, FR2006}
\begin{eqnarray}
	(\tilde{\mathcal{L}}_{\Delta})_{\mathbf{k}\mathbf{q}} = e^{-\Delta \mathbf{q}^2}\int_{\mathbb{T}^2}e^{2\pi \mathrm{i}(\mathbf{q}\cdot \mathbf{x} - \mathbf{k} \cdot f(\mathbf{x}))}\mathrm{d}\mathbf{x} \,.
	\label{eq:fp_fourier}
\end{eqnarray}
Note that the Perron-Frobenius operator $\mathcal{L}$ for area-preserving maps acting on $L^2$ is unitary. This implies that the sub-unit eigenvalues of the Perron-Frobenius operator are all zero since the spectrum lies entirely on the unit circle of the complex plane. On the other hand, Faure and Roy proved that, for a class of nonlinear uniformly hyperbolic 2-dimensional maps, which includes $C$ and $C_{\epsilon,\nu}$ with small enough $\epsilon$, the eigenvalues of the Fokker-Planck operator $\tilde{\mathcal{L}}_{\Delta}$ tend to the \textit{Ruelle-Pollicott resonances} of the Perron-Frobenius operator as $\Delta \rightarrow 0$ \cite{FR2006}. 

The Ruelle-Pollicott resonances are alternatively obtained from a conjugated operator to the Perron-Frobenius, acting on a space of regular functions. If the result in \cite{FR2006} extends to the map $C_{\epsilon,\nu}$ with a larger perturbation, we may identify the numerically obtained spectra of $\tilde{\mathcal{L}}_{\Delta}$ with the Ruelle-Pollicott resonances, in the noiseless limit. Since we are not aware of any proof of that, we refrain from using the Ruelle-Pollicott terminology in what follows.

\subsection{Unperturbed cat map}
For the cat map $C$, we can easily show the sub-unit eigenvalues of the Fokker-Planck operator are all zero for any $\Delta$. Since the matrix elements of $\tilde{\mathcal{L}}_{\Delta}$ can be explicitly given as \cite{FR2006},
\begin{eqnarray}
	(\tilde{\mathcal{L}}_{\Delta})_{\mathbf{k}\mathbf{q}} = e^{-\Delta \mathbf{q}^2}\delta_{\mathbf{k}\cdot \bar{C}, \mathbf{q}}, 
\end{eqnarray}
and $\tilde{\mathcal{L}}_{\Delta}$ only has one nonzero diagonal element $(\tilde{\mathcal{L}}_{\Delta})_{\mathbf{0}\mathbf{0}}$, while all other diagonal entries are equal to 0. Here $\bar{C}= \displaystyle \left( \begin{array}{l} 1 \quad 1 \\ 1 \quad 2 \end{array} \right)$ and $\mathbf{0}=(0,0)$. Thus, the sub-unit eigenvalues of the Fokker-Planck operator are all zero. The same result can be obtained for the Perron-Frobenius operator \cite{San2002}. We therefore only consider the perturbed cat map for the analysis of the second eigenvalues and eigenfunctions of the Fokker-Planck operator.

\subsection{Perturbed cat map}
For the perturbed cat map $C_{\epsilon, \nu}$, the integral (\ref{eq:fp_fourier}) is evaluated in terms of the Bessel function of the first kind $I_{\omega}$, as follows \cite{TC2003}:
\begin{eqnarray}
	(\tilde{\mathcal{L}}_{\Delta})_{\mathbf{k}\mathbf{q}} =
	\left\{
	\begin{array}{l}
		e^{-\Delta \mathbf{q}^2} \delta_{0, -Q_x}(-1)^{\frac{Q_y}{\nu}}I_{-\frac{Q_y}{\nu}} \left( (k_x + k_y)2\pi \frac{\epsilon}{\nu} \right) \quad \; \frac{Q_y}{\nu} \in \mathbb{Z}, \\
		0  \qquad \qquad \qquad \qquad \qquad \qquad \qquad \qquad \mathrm{otherwise}.
	\end{array}
	\right.
\end{eqnarray}
where $(Q_x, Q_y) = \mathbf{k}\cdot C - \mathbf{q}$. For numerical calculations, we truncate the wavenumber, so that $\mathbf{k},\mathbf{q} \in [-K, K]\times[-K,K],\, K > 0$, and the Fokker-Planck operator $\tilde{\mathcal{L}}_{\Delta}$ is expressed as a $(2K + 1)^2 \times (2K + 1)^2$-dimensional matrix.

In the following calculation, we take $\nu = 1$ because the second-largest eigenvalue is real-valued, isolated, and non-degenerate in magnitude, as explained below. On the other hand, the number of second-largest eigenvalues increases for $\nu \ge 2$ and complex-valued second-largest eigenvalues appear, making the overall picture more complicated. This aspect will be closely examined in the next section.

\subsubsection{Convergence of second-largest eigenvalues}
\begin{figure}[H]
	\subfigure[\hspace{4cm}]{
		\begin{minipage}{0.5\hsize}
			\includegraphics[width = \hsize, bb = 0 0 460.8 345.6]{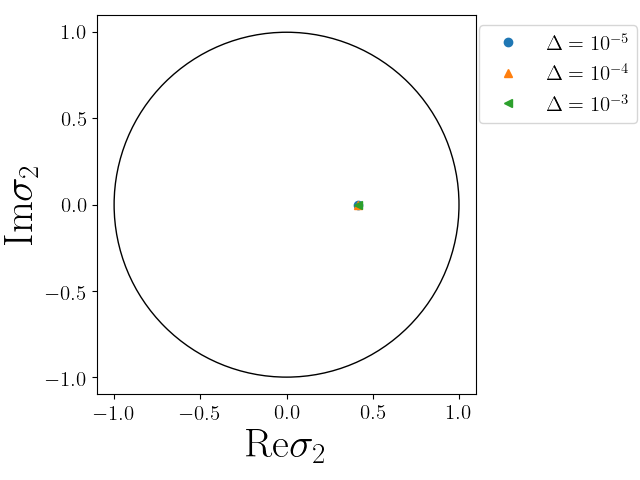}
		\end{minipage}
	}
	\subfigure[\hspace{4cm}]{
		\begin{minipage}{0.5\hsize}
			\includegraphics[width = \hsize, bb = 0 0 460.8 345.6]{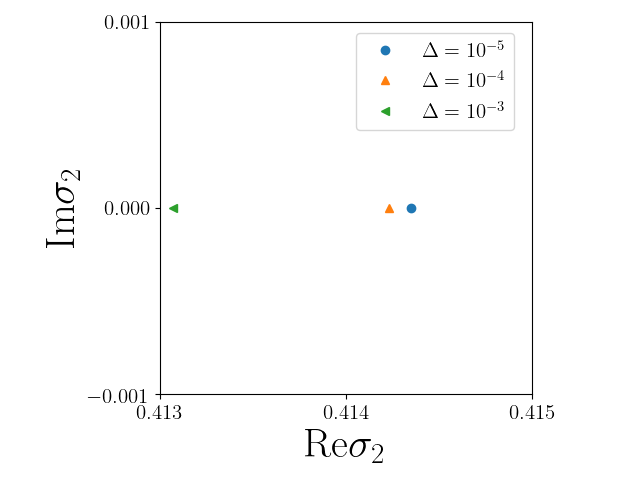}
		\end{minipage}
	}
	\caption{
		The second-largest eigenvalue of the Fokker-Planck operator for the perturbed cat map $C_{\epsilon,\nu}$. We take $\epsilon = 0.1, \nu = 1$ and $K=100$. The circle represents the unit circle. (b) is the magnification of (a).
	}
	\label{fig:fp_sle_on_plane_nu1}
\end{figure}

We first provide numerical evidence that the Fokker-Planck operator for the perturbed cat map has non-zero second-largest eigenvalues. In figure \ref{fig:fp_sle_on_plane_nu1}, we give the location of the second-largest eigenvalues for $\epsilon = 0.1$. It is clearly seen that with decrease in diffusivity $\Delta$ the location of the second-largest eigenvalue gets stabilized and tends to be fixed. In this calculation, the maximum wavenumber is taken to be $K=100$.	As shown in table \ref{tab:epsilon_vs_sigma2}, the second-largest eigenvalue gradually decreases with decrease in the perturbation $\epsilon$. This behavior implies that all sub-unit eigenvalues tend to 0 as $\epsilon \to 0$, which is consistent with the mathematical result for the cat map \cite{FR2006}. 
\begin{table}[htb]
		\begin{center}
		\begin{tabular}{|c|c|} \hline
			Perturbation $\epsilon$ & Second-largest eigenvalue $\sigma_2$  \\ \hline \hline
			$0.12$ & $0.5284306\dots$ \\ \hline
			$0.1$ & $0.4130742\dots$ \\ \hline
			$0.08$ & $0.3123028\dots$ \\ \hline
			$0.06$ & $0.2220071\dots$ \\ \hline
			$0.04$ & $0.1403190\dots$ \\ \hline
		\end{tabular}
		\caption{The second-largest eigenvalue of the Fokker-Planck operator with different perturbation $\epsilon$. We take $K = 100$ and $\Delta = 10^{-3}$.}
		\label{tab:epsilon_vs_sigma2}
	\end{center}
\end{table}

\begin{figure}[H]
	\subfigure[\hspace{4cm}]{
		\begin{minipage}{0.5\hsize}
			\begin{center}
				\includegraphics[width = \hsize, bb=0 0 460.8 345.6]{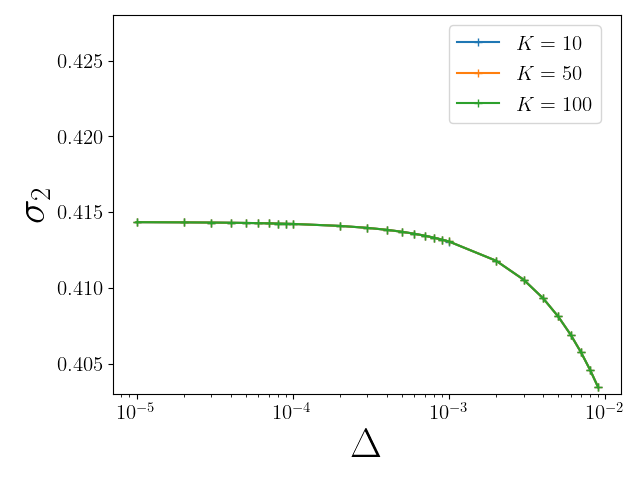}
			\end{center}
		\end{minipage}
	}
	\subfigure[\hspace{4cm}]{
		\begin{minipage}{0.5\hsize}
			\begin{center}
				\includegraphics[width = \hsize, bb=0 0 460.8 345.6]{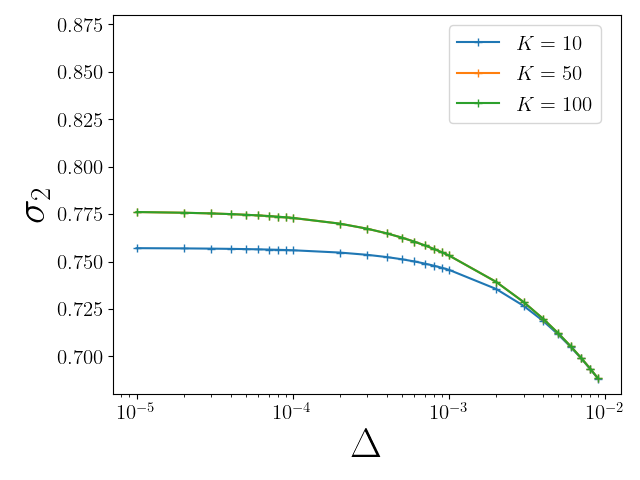}
			\end{center}
		\end{minipage}
	}
	\caption{
		Absolute value of the second-largest eigenvalue of the Fokker-Planck operator for the perturbed cat map vs the diffusivity $\Delta$. (a) $\epsilon = 0.1$ and (b) $\epsilon = 0.15$.}
	\label{fig:fp_sle_vs_delta_nu1}
\end{figure}

We can more clearly show that the second-largest eigenvalues actually converge to certain non-zero constant values, by plotting the position of eigenvalues as a function of $\Delta$. Figure \ref{fig:fp_sle_vs_delta_nu1} illustrates these results for several values of the maximum wavenumber $K$. For $\epsilon = 0.1$, the curves relative to different $K$ are indistinguishable, as they all collapse to a single curve, and they all tend to the same value for $\Delta \rightarrow 0$. It is noted that the convergence is stable compared to the result obtained using the Ulam method.

On the other hand, for $\epsilon = 0.15$, the cutoff $K=10$ is not enough to achieve well-convergent result, but for larger $K$ values, we find that $|\sigma_2|$ seems to  stabilize on a limit value for decreasing diffusivity, implying that the second-largest eigenvalue takes a non-zero value in this case as well in the noiseless limit. We have verified that all these results do not change even if we replace the numerical precision from double to quadruple. 

\begin{figure}[H]
	\begin{center}
		\begin{minipage}{0.5\hsize}
			\includegraphics[width = \hsize, bb=0 0 400 330]{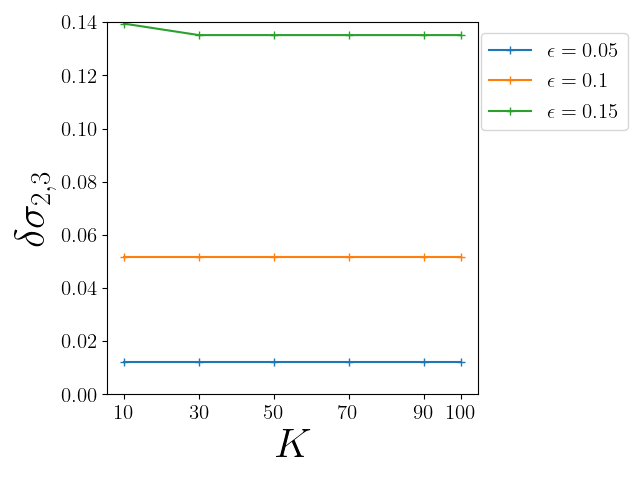}
		\end{minipage}
	\end{center}
	\caption{The difference between the absolute values of the second-largest eigenvalue and third-largest eigenvalue as a function of $K$. We take $\Delta = 10^{-3}$.}
	\label{fig:fp_diff_sle_tle_nu1}
\end{figure}

We can further confirm that the second-largest eigenvalue thus obtained is isolated from the third-largest and other eigenvalues as noticed in figure \ref{fig:fp_diff_sle_tle_nu1}. Here we plot the distance $\delta \sigma_{2,3}:= \vert \sigma_2 \vert - \vert \sigma_3 \vert$, where $\sigma_2$ and $\sigma_3$ denote the second and third-largest eigenvalue for the Fokker-Planck operator, respectively. The third-largest eigenvalue is also real-valued. The distance $\delta \sigma_{2,3}$ again tends to a fixed value, meaning that there exists a finite gap between the second-largest and other eigenvalues and so the second-largest one is isolated.

These numerical calculations altogether strongly suggest that the second-largest eigenvalue of the Fokker-Planck operator is real valued, strictly positive, and isolated, which is in contrast with what seen for the linear cat map $C$. When developing a rigorous argument, one must treat the order of the limit $\Delta \rightarrow 0$ and $K \rightarrow \infty$ carefully.

\subsubsection{Spatial patterns of second eigenfunctions}
\begin{figure}[H]
	\subfigure[\hspace{4cm}]{
		\begin{minipage}{0.5\hsize}
			\begin{center}
				\includegraphics[width = \hsize, bb=0 0 460.8 345.6]{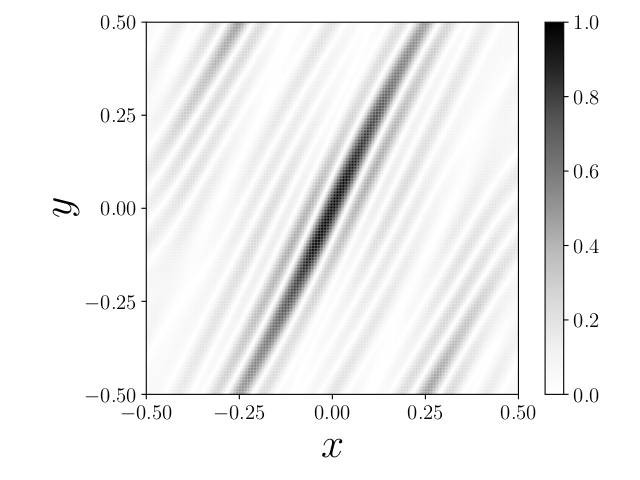}
			\end{center}
		\end{minipage}
	}
	\subfigure[\hspace{4cm}]{
		\begin{minipage}{0.5\hsize}
			\begin{center}
				\includegraphics[width = \hsize, bb=0 0 460.8 345.6]{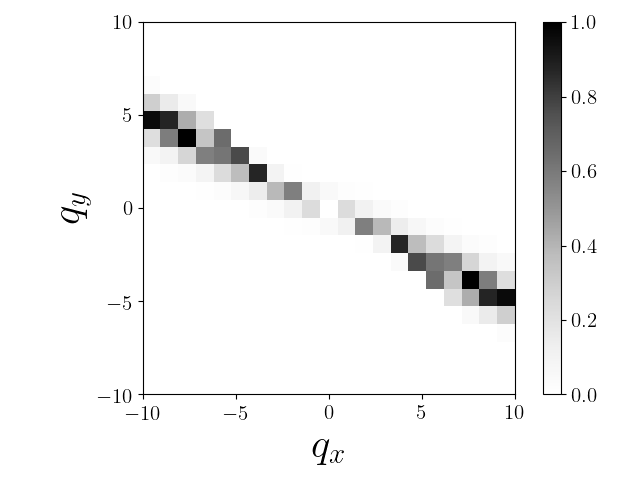}
			\end{center}
		\end{minipage}
	}
	\subfigure[\hspace{4cm}]{
		\begin{minipage}{0.5\hsize}
			\begin{center}
				\includegraphics[width = \hsize, bb=0 0 460.8 345.6]{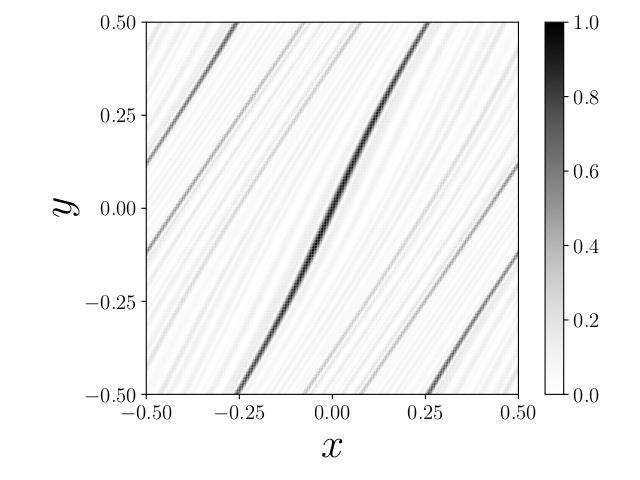}
			\end{center}
		\end{minipage}
	}
	\subfigure[\hspace{4cm}]{
		\begin{minipage}{0.5\hsize}
			\begin{center}
				\includegraphics[width = \hsize, bb=0 0 460.8 345.6]{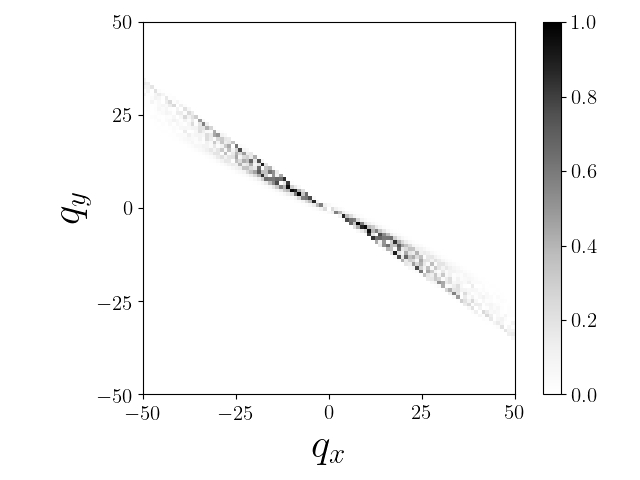}
			\end{center}
		\end{minipage}
	}
	\subfigure[\hspace{4cm}]{
		\begin{minipage}{0.5\hsize}
			\begin{center}
				\includegraphics[width = \hsize, bb=0 0 460.8 345.6]{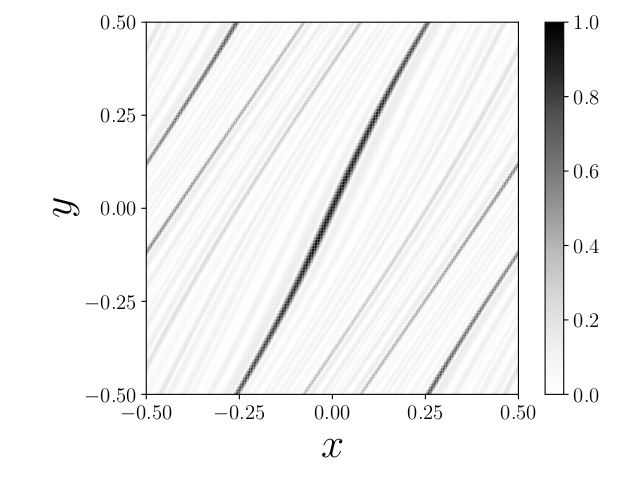}
			\end{center}
		\end{minipage}
	}
	\subfigure[\hspace{4cm}]{
		\begin{minipage}{0.5\hsize}
			\begin{center}
				\includegraphics[width = \hsize, bb=0 0 460.8 345.6]{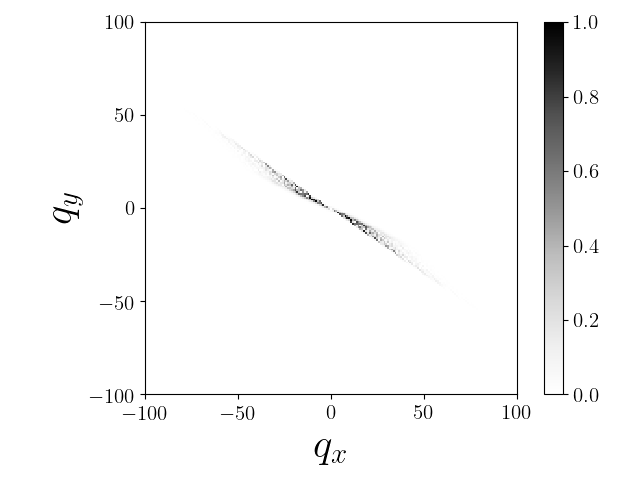}
			\end{center}
		\end{minipage}
	}
	\caption{
		Magnitude of the second eigenfunctions of the Fokker-Planck operator for the perturbed cat map (a)-(b) $K = 10$, (c)-(d) $K = 50$ and (e)-(f) $K = 100$. Left panels: phase space representation. Right panels: wavenumber representation. $\epsilon = 0.1$, $\nu = 1$ and $\Delta = 5 \times 10^{-3}$ are taken. 
	}
	\label{fig:fp_second_eigenfunc_nu1}
\end{figure}

On the basis of the well convergent nature of eigenvalues confirmed above, we might expect that the associated second eigenfunctions can also be obtained via the Fokker-Planck operator. As is shown in figure \ref{fig:fp_second_eigenfunc_nu1}, it is indeed the case. We explain this by focusing on the $K$-dependence with $\Delta$ being fixed. As is seen at the first column of figure \ref{fig:fp_second_eigenfunc_nu1}, the pattern becomes stabilized as the maximum wavenumber increases. Since we may expect that the smaller $\Delta$ is, the finer spatial structure is developed, there should exist a minimum cutoff $K$ such that the spatial pattern produced by the Fokker-Planck operator with diffusivity $\Delta$ could be resolved. The convergence of the observed patterns could be achieved because such a desired condition is satisfied in this case. 

In order to estimate the maximum wavenumber necessary to provide well-convergent eigenfunctions, we present eigenfunctions in the wavenumber representation. As shown in figures~\ref{fig:fp_second_eigenfunc_nu1} (b),(d) and (f), the distribution in the wave number space exceeds the wave range if $K=10$ is taken. However, the distribution is reasonably confined in the range with $K=50$, and well confined in the case of $K=100$. These observations are consistent with the phase space representations shown in the left panels in figure~\ref{fig:fp_second_eigenfunc_nu1} the spatial patterns for the cases with $K=50$ and $100$ are almost the same, suggesting that these results are convergent. 

\begin{figure}[H]
	\subfigure[\hspace{4cm}]{
		\begin{minipage}{0.5\hsize}
			\begin{center}
				\includegraphics[width = \hsize, bb=0 0 460.8 345.6]{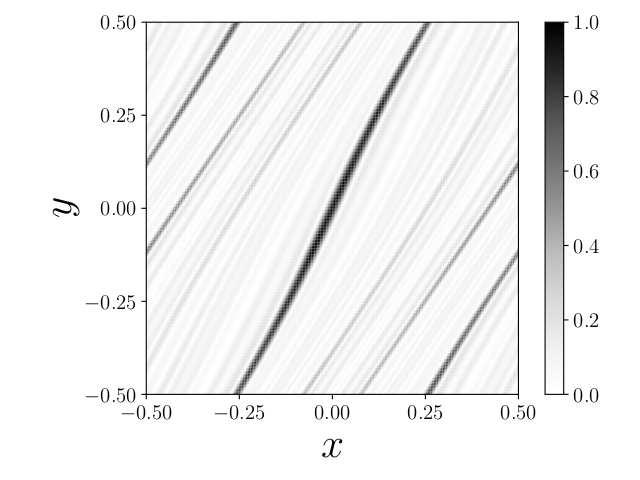}
			\end{center}
		\end{minipage}
	}
	\subfigure[\hspace{4cm}]{
		\begin{minipage}{0.5\hsize}
			\begin{center}
				\includegraphics[width = \hsize, bb=0 0 460.8 345.6]{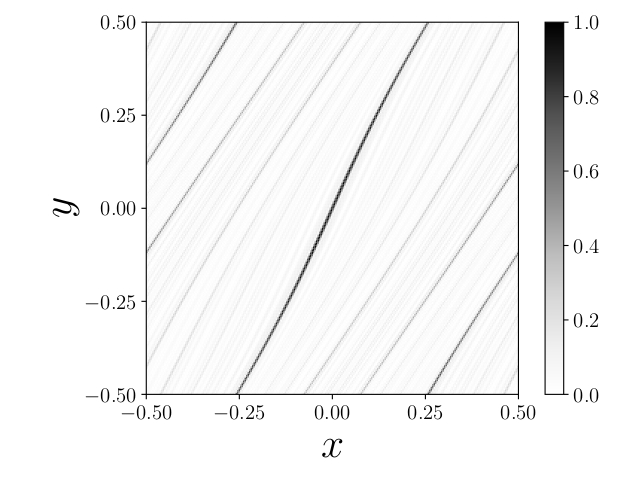}
			\end{center}
		\end{minipage}
	}
	\caption{
			The second eigenfunctions of the Fokker-Planck operator for the perturbed cat map (a) $\Delta =9 \times 10^{-3}$ and (b) $\Delta = 10^{-3}$. We take $\epsilon = 0.1$, $\nu = 1$ and $K = 100$.  
	}
	\label{fig:fp_second_eigenfunc_nu1_K100}
\end{figure}

As shown in figure \ref{fig:fp_second_eigenfunc_nu1_K100}, the localized region of the second eigenfunction becomes narrower with decrease of the diffusivity $\Delta$. From this observation, we would expect that the spatial pattern tends to a fractal set as $\Delta$ goes to zero, while maintaining the spatial inhomogeneity that surrounds the fixed point. 

From this convergent nature in the second-largest eigenvalues and the corresponding eigenfunctions for the Fokker-Planck operator, we might expect that the second-largest eigenvalue for the Perron-Frobenius operator takes a finite value, and observed spatial inhomogeneity survives even in the limit of $\Delta \to 0$. The small diffusivity limit is highly nontrivial and subtle, in particular for the eigenfunctions since they must become a class of hyperfunctions. The eigenfunction for the quadbaker map, which was rigorously obtained through the Ulam method, could be an analogous example showing an inhomogeneous spatial profile \cite{Fro2007}.  

Our numerical calculations strongly suggest that the second-largest eigenvalue of the Fokker-Planck operator of the perturbed cat map tends to a finite positive value as $\Delta \rightarrow 0$, and the second eigenfunction exhibits localization similarly observed in the one obtained using the Ulam method (cf. figure \ref{fig:ulam_perturbed_cat_second_eigenvec}). In the subsequent sections, we will explore the origin of localization and what quantity characterizes localization patterns.

%% file: 5_inhomogeneity_unst_mani.tex
It is worth noting that, as was observed in spatial patterns obtained via the Ulam method, strongly localized regions appear in the eigenfunctions of the evolution operator. This tells us that the second eigenfunction, which provides information on the decay of correlations in the dynamics, is spatially inhomogeneous. Such a signature reminds us of the position dependence of the escape rate, a subject that has recently attracted much attention \cite{KL2009, BY2011, Bun2012, APT2013}.

\begin{figure}[H]
	\subfigure[\hspace{4cm}]{ 
		\begin{minipage}{0.5\hsize}
			\begin{center}
				\includegraphics[width = \hsize, bb=0 0 460.8 345.6]{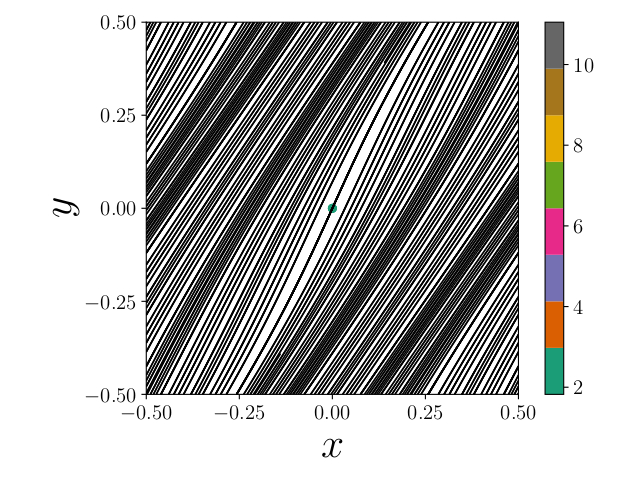}
			\end{center}
		\end{minipage}
	}
	\subfigure[\hspace{4cm}]{ 
		\begin{minipage}{0.5\hsize}
			\begin{center}
				\includegraphics[width = \hsize, bb=0 0 460.8 345.6]{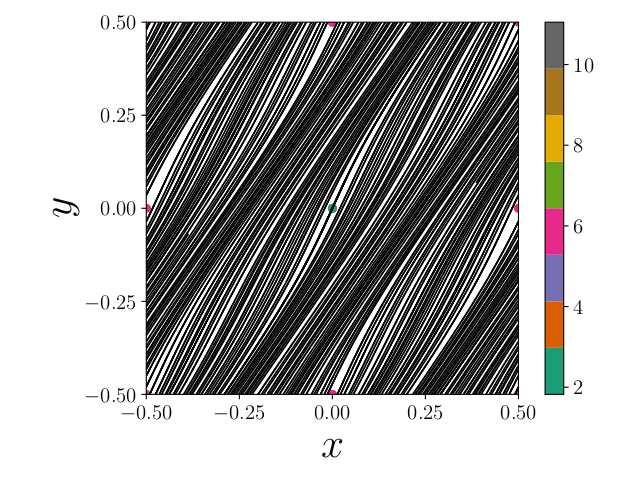}
			\end{center}
		\end{minipage}
	}
	\subfigure[\hspace{4cm}]{ 
		\begin{minipage}{0.5\hsize}
			\begin{center}
				\includegraphics[width = \hsize, bb=0 0 460.8 345.6]{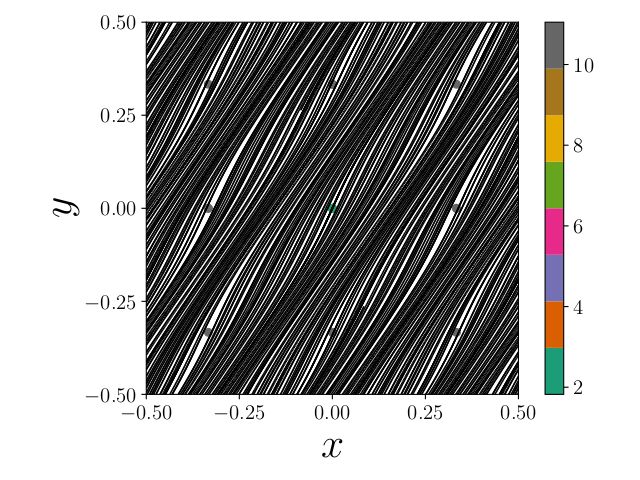}
			\end{center}
		\end{minipage}
	}
	\subfigure[\hspace{4cm}]{ 
		\begin{minipage}{0.5\hsize}
			\begin{center}
				\includegraphics[width = \hsize, bb=0 0 460.8 345.6]{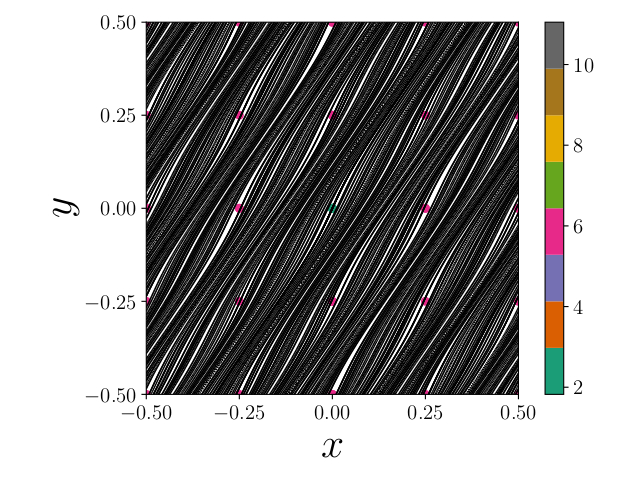}
			\end{center}
		\end{minipage}
	}
	\caption{
		Solid line: unstable manifold of the perturbed cat map for (a) $\nu = 1$, (b) $\nu = 2$, (c) $\nu = 3$, and (d) $\nu = 4$. $\epsilon = 0.1$ is taken. Dots: unstable periodic orbits with a relatively small magnitude of the local instability (see equation~(\ref{eq:instability_pcm})). Color code: stability multipliers of the periodic orbits.
	}
	\label{fig:unstable_mani_pcm}
\end{figure}

Here we just point out that the localized region in the eigenfunction patterns also features a sparse unstable manifold. As displayed in figure \ref{fig:unstable_mani_pcm}(a), the density of an unstable manifold emanating from a fixed point is not uniform but inhomogeneous. Sparse regions seem to coincide with localized regions of the second eigenfunction of the Fokker-Planck operator (cf. figure~\ref{fig:fp_second_eigenfunc_nu1}) and the Ulam matrix (cf. figure~\ref{fig:ulam_perturbed_cat_second_eigenvec}). Spatial profiles of the second eigenfunction of the Fokker-Placnk operator for smaller diffusivity case is close to those obtained based on the Ulam method. 

In what follows, we show that unstable manifolds are sparser in proximity of short periodic orbits with small stability exponents. Recall that the larger eigenvalue of the Jacobian matrix of the map is here referred to as the stability multiplier. For the perturbed cat map $C_{\epsilon, \nu}$ it is given as 
\begin{eqnarray}
	\lambda_{\epsilon, \nu}^{(P)}(y) = \frac{-(2\pi \epsilon \cos (2\nu \pi y) - 3) + \sqrt{(2\pi \epsilon \cos (2\nu \pi y)- 3)^2 - 4}}{2},
	\label{eq:instability_pcm}
\end{eqnarray}
which takes the minimum value on the line $\nu y \in \mathbb{Z}$ provided that $0 < \epsilon < 1/2\pi$ is satisfied. One can verify that the set of periodic orbits $\{ (x, y) = (p / \nu - \lfloor p/\nu + 1/2 \rfloor, q / \nu - \lfloor q/\nu + 1/2 \rfloor) ; p, q = 0, 1, \dots, \nu - 1\}$ has minimum stability multiplier.

As demonstrated in figure \ref{fig:unstable_mani_pcm}, the region where the unstable manifold is sparsely running is populated with relatively shorter, and less unstable periodic orbits. Here the stability is measured by the local instability, defined by $\vert \lambda_{\epsilon, \nu}^{(P)} \vert ^p$, where $p$ denotes the period of the periodic orbit. We here present not only the $\nu = 1$ case but also the $\nu \ge 2$ cases. This coincidence could simply be understood by considering the fact that trajectories initially in the vicinity of a short, lesser unstable periodic orbit, take a longer time to escape from the initial region, implying that the inverse process of entering a less unstable region also takes a longer time to occur, which gives rise to the sparse densities of the unstable manifold. 

From this observation, one may reduce the origin of localization of the second eigenfunctions to the difference of local stability of the dynamics. However, as clarified in the next section, that explanation is simplistic all by itself, although local instabilities have been used to understand quantum scarring \cite{KapHel98}. 

%% file: 6_finitetimelyapunov.tex
In this section we explore what signatures of dynamics are imprinted in the spatial patterns of the second eigenfunction. As shown in the previous section, spatial patterns of the second eigenfunction well reflect the density or sparseness of the unstable manifold. In particular, it was observed that the second eigenfunction is localized around the relatively lesser unstable periodic orbits.

\begin{figure}[H]
	\begin{center}
		\begin{minipage}{0.5\hsize}
			\begin{center}
				\includegraphics[width = \hsize, bb= 0 0 460.8 345.6]{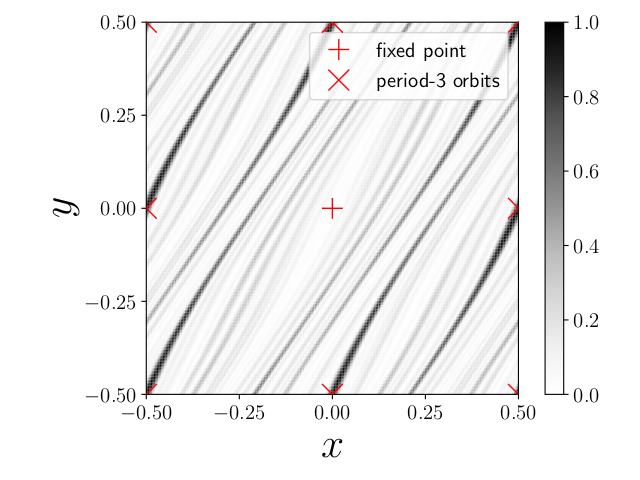}
			\end{center}
		\end{minipage}
	\end{center}
	\caption{
		The absolute value of the second eigenfunction of the Fokker-Planck operator associated with the perturbed cat map for $\epsilon = 0.1$ and $\nu = 2$. Note that the corresponding eigenvalue takes a complex value (see below).  We take $K = 50$ and $\Delta = 10^{-3}$. The fixed point ($+$) and period-3 orbit ($\times$) are marked in red. 
	}
	\label{fig:second_eigenfunc_localize_period3}
\end{figure}

We here present an example demonstrating that, unlike in the argument developed in quantum scarring, the local instability of the periodic orbits does not necessarily control the localization observed in the second eigenfunction of the Perron-Frobenius operator. Figure \ref{fig:second_eigenfunc_localize_period3} portrays a second eigenfunction in the case of the perturbed cat map with $\nu = 2$. As noticed, this second eigenfunction is not localized around the periodic orbit with the smallest instability, that is the fixed point at the origin, but rather around the period-3 periodic orbit. We will further discuss how this second eigenfunction behaves when an additional perturbation is applied.

Here we consider the maximal finite-time (mFT) Lyapunov exponent as a possible candidate to understand localization. The mFT Lyapunov exponent is a characteristic quantity defined at each point in phase space, but it carries information about a longer time scale than that of linearized dynamics. 

\subsection{Distribution of the maximum finite-time Lyapunov exponent}
\label{sec:mFTLEdist}
\begin{figure}[H]
	\subfigure[\hspace{4cm}]{
		\begin{minipage}{0.5\hsize}
			\begin{center}
				\includegraphics[width = \hsize, bb = 0 0 460.8 345.6]{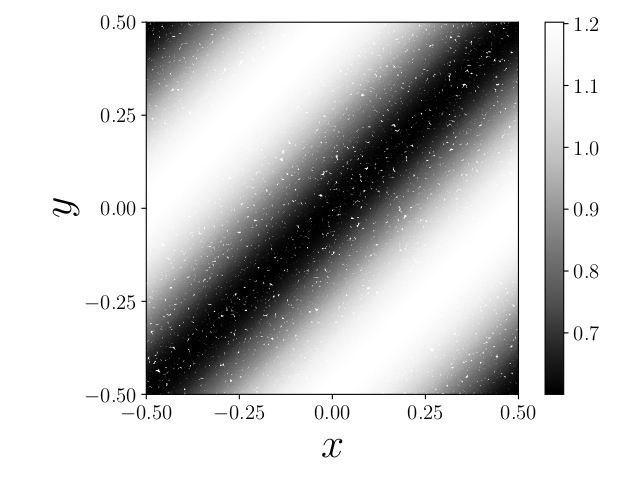}
			\end{center}
		\end{minipage}
	}
	\subfigure[\hspace{4cm}]{
		\begin{minipage}{0.5\hsize}
			\begin{center}
				\includegraphics[width = \hsize, bb = 0 0 460.8 345.6]{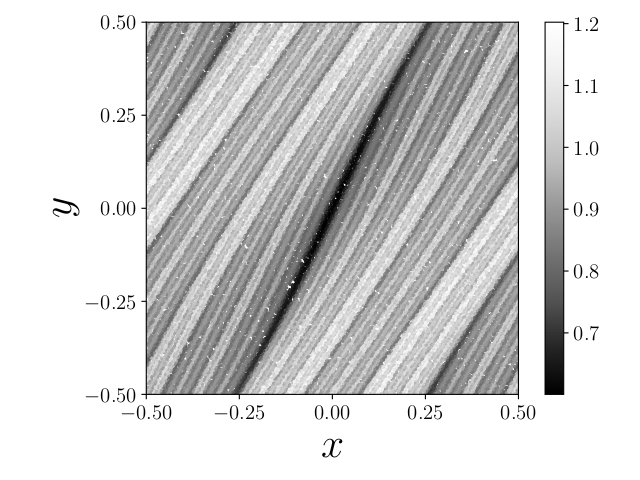}
			\end{center}
		\end{minipage}
	}
	\subfigure[\hspace{4cm}]{
		\begin{minipage}{0.5\hsize}
			\begin{center}
				\includegraphics[width = \hsize, bb = 0 0 460.8 345.6]{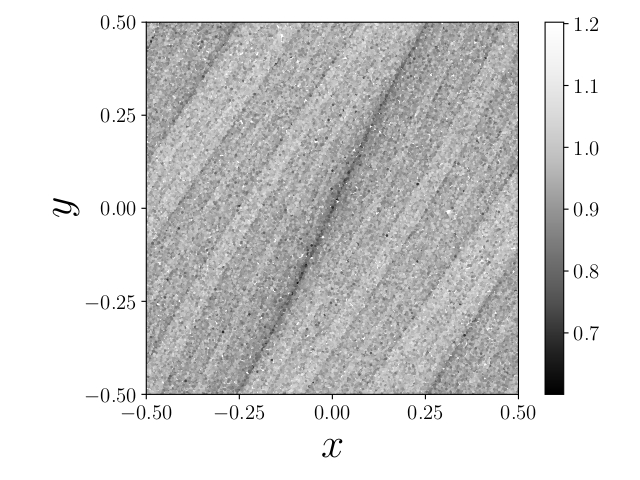}
			\end{center}
		\end{minipage}
	}
	\subfigure[\hspace{4cm}]{ 
		\begin{minipage}{0.5\hsize}
			\begin{center}
				\includegraphics[width = \hsize, bb=0 0 460.8 345.6]{figure14_e.jpg}
			\end{center}
		\end{minipage}
	}
	\caption{
		(a)-(c) Distribution of the mFT Lyapunov exponent of the perturbed cat map with $\epsilon = 0.1$, $\nu=1$.  The time interval to evaluate the mFT Lyapunov exponent is 	(a) $t=1$, (b) $t=5$ and (c) $t=15$. $10^5$ initial points randomly distributed in the entire phase space are taken for calculations. (d) The second eigenfunction of the Fokker-Planck operator of the perturbed cat map for $\epsilon = 0.1$ and $\nu = 1$. We take $K=100$ and $\Delta = 5 \times 10^{-3}$.
	}
	\label{fig:lyapunov_nu1}
\end{figure}

We first introduce the mFT Lyapunov exponent. To this end, define the matrix $A$ as, 
\begin{eqnarray}
    A(\mathbf{x}_0, t) = \prod_{k = 0}^{t - 1}J(\mathbf{x}_k),
		\qquad  J_{ij}(\mathbf{x}) = \frac{\partial f_i}{\partial x_j} \, , 
\end{eqnarray}
where $\mathbf{x}_k = f(\mathbf{x}_{k - 1})$ and $J(\mathbf{x})$ is the Jacobi matrix of $f$ at $\mathbf{x}$. Let $a(\mathbf{x}_0, t)$ be the larger eigenvalue of $A(\mathbf{x}_0,t)$ in magnitude. The mFT Lyapunov exponent at time $t$ is expressed as, 
\begin{eqnarray}
    \Lambda(\mathbf{x}_0,t) = \frac{1}{t}\mathrm{ln}\vert a(\mathbf{x}_0,t)\vert.
\end{eqnarray}
In figure \ref{fig:lyapunov_nu1}, we show the distribution of the mFT Lyapunov exponent for the perturbed cat map as a function of $\mathbf{x}_t = (x_t, y_t)$. We take the perturbation strength $\epsilon = 0.1$ and the perturbation frequency $\nu = 1$. For $t=1$, the mFT Lyapunov exponent is just the logarithm of the larger eigenvalue of the Jacobian matrix. As shown in figure~\ref{fig:lyapunov_nu1} (a), the distribution does not reflect structures of the unstable manifold, but only the linearized dynamics, therefore what is imprinted in the localized pattern of eigenfunctions is not information associated with local stability multipliers. For $t>1$, as shown in figure \ref{fig:lyapunov_nu1} (b) and (c), the distribution of the mFT Lyapunov exponents is smooth along the unstable manifold and wildly oscillating along the stable manifold. As time proceeds, the distribution develops a finer structure, and the mFT Lyapunov exponent takes smaller values inside sparse regions of the unstable manifold. As mentioned in the previous section, the second eigenfunction of the Fokker-Planck operator is localized in the sparse regions of the unstable manifold. We expect that the mFT Lyapunov exponent reflects the pattern of the second eigenfunction of the Fokker-Planck operator. This is indeed the case, as shown in figure~\ref{fig:lyapunov_nu1}(b)-(c).

Note that the mFT Lyapunov exponent here is concerned with the noiseless system while the Fokker-Planck operator is the evolution operator for the noisy system. However, as argued in section \ref{sec:fokkerplanck}, the second eigenfunction of the Fokker-Planck operator keeps spatial inhomogeneity with decrease of the diffusivity $\Delta$, although the limiting distribution will be a non-smooth function. In this section, we therefore identify the patterns of localization of the second eigenfunction of the Fokker-Planck operator with sufficiently small $\Delta$ with those of the analogous eigenfunction of the noiseless (Perron-Frobenius) operator.

\subsection{mFT Lyapunov exponent distribution and second eigenfunction}
We here consider the relation between the second eigenfunction of the Perron-Frobenius operator and the distribution of the mFT Lyapunov exponents by assuming that the Perron-Frobenius operator for a 2-dimensional uniformly hyperbolic area-preserving map $f:\mathcal{M}\rightarrow \mathcal{M}$ has a real isolated second-largest eigenvalue $\sigma$ of eigenfunction $\psi$. 

\begin{figure}[H]
	\subfigure[\hspace{4cm}]{
		\begin{minipage}{0.5\hsize}
			\begin{center}
				\includegraphics[width = \hsize, bb = 0 0 460.8 345.6]{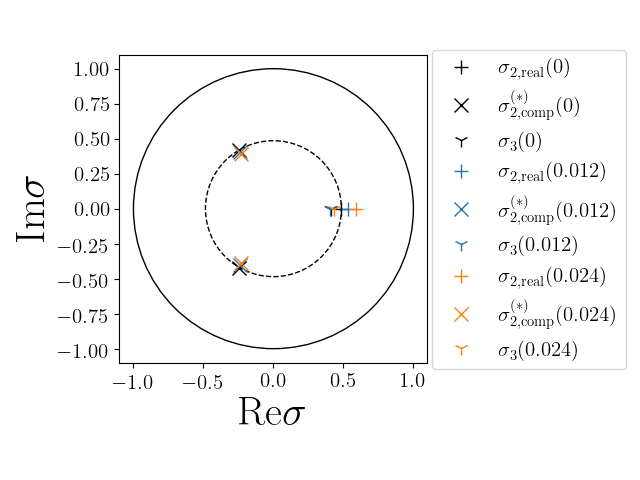}
			\end{center}
		\end{minipage}
	}
	\subfigure[\hspace{4cm}]{
		\begin{minipage}{0.5\hsize}
			\begin{center}
				\includegraphics[width = \hsize, bb = 0 0 460.8 345.6]{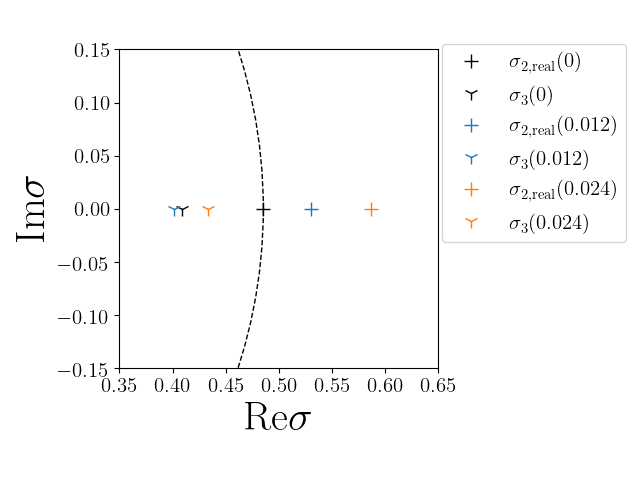}
			\end{center}
		\end{minipage}
	}
	\subfigure[\hspace{4cm}]{
		\begin{minipage}{0.5\hsize}
			\begin{center}
				\includegraphics[width = \hsize, bb = 0 0 460.8 345.6]{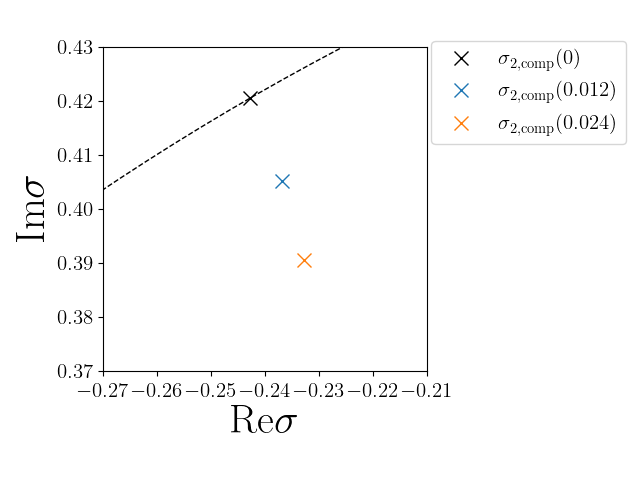}
			\end{center}
		\end{minipage}
	}
	\subfigure[\hspace{4cm}]{
		\begin{minipage}{0.5\hsize}
			\begin{center}
				\includegraphics[width = \hsize, bb = 0 0 460.8 345.6]{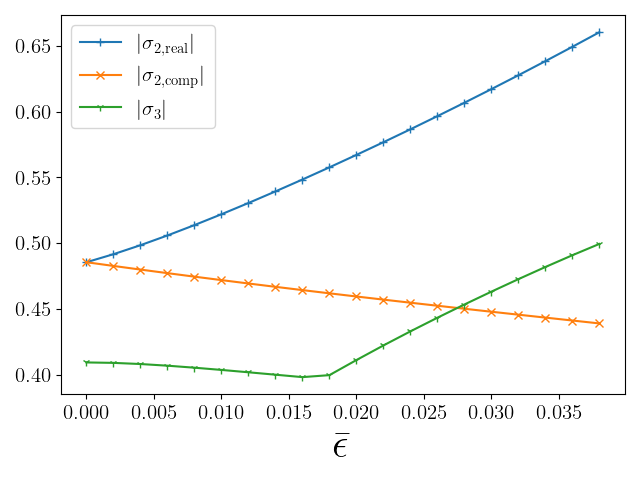}
			\end{center}
		\end{minipage}
	}
	\caption{
	(a)-(c) Eigenvalues $\sigma_{2, \mathrm{real}}(\bar{\epsilon}), \sigma_{2, \mathrm{comp}}(\bar{\epsilon}), \sigma_{2,\mathrm{comp}}^{*}(\bar{\epsilon})$ and $\sigma_{3}(\bar{\epsilon})$ of the Fokker-Planck operator for the map ${\cal C}_{\epsilon,\bar{\epsilon}}$ with $\epsilon = 0.1$. We take $K=30$ and $\Delta = 10^{-3}$. (b) A magnification of (a) around $\sigma_{2, \mathrm{real}}$ and $\sigma_{3}$. (c) A magnification of (a) around $\sigma_{2. \mathrm{comp}}$. (d) Absolute values $\vert \sigma_{2, \mathrm{real}}(\bar{\epsilon})\vert , \vert \sigma_{2, \mathrm{comp}}(\bar{\epsilon})\vert$ and $\vert \sigma_{3}(\bar{\epsilon})\vert $ as a function of $\bar{\epsilon}$. 
	}
	\label{fig:double_pcm_eigenval}
\end{figure}

For this purpose, let us consider the time evolution of a distribution function $\rho(\mathbf{x})$, which is initially smooth in $\mathcal{M}$ and takes positive values. In the long time limit, $\rho(\mathbf{x})$ tends to a uniform distribution (\textit{cf.} \S \ref{sec:perronfrobenius}), which we normalize to unity in what follows, and the relaxation toward the equilibrium distribution is expected to behave as 
\begin{eqnarray}
	\mathcal{L}^t \rho(\mathbf{x}) \sim 1 + c \sigma^t \psi(\mathbf{x}),
	~~~~~( t \to \infty)
	\label{eq:pf_approx}
\end{eqnarray}
where $c \in \mathbb{R}$ is some constant. Note that the convergence to the equilibrium state occurs not in a point-wise manner but in the sense of weak convergence at most \cite{Mac1992}. From equation (\ref{eq:pf_approx}), we infer that a proper time for evaluating the mFT Lyapunov exponent is $t \sim -1/\log|\sigma|$.  The distributions of figure \ref{fig:lyapunov_nu1}, and the eigenvalues of the Fokker-Planck operator reported in  figure \ref{fig:double_pcm_eigenval} enable us to verify this estimate.

We introduce a coordinate $\xi \in (-\infty, \infty)$ along the unstable manifold in $\mathcal{M}$, and consider the dynamics $f_u$ restricted on the unstable manifold. The Perron-Frobenius operator associated with $f_u$ is expressed as
\begin{eqnarray}
	\mathcal{L}_u \rho _u(\xi) &= \int_{\mathcal{M}} \delta(\xi - f_u(\bar{\xi})) \rho_u(\bar{\xi}) \mathrm{d}\bar{\xi} \nonumber \\
	&=\int_{\mathcal{M}} \delta(\xi - \eta ) \rho_u(f_u^{-1}(\eta)) \vert  f_u'(\eta)\vert ^{-1} \mathrm{d}\eta \nonumber \\
	&=\vert f_u'(\xi) \vert ^{-1}\rho_u(f_u^{-1}(\xi)), 
	\nonumber 
\end{eqnarray}
so the $t$-time iteration of the Perron-Frobenius operator is
\begin{eqnarray}
	\mathcal{L}_u^{t} \rho _u(f^t(\xi)) = \prod_{k = 0}^{t - 1} \left\vert  f_u' (f_u^k(\xi)) \right\vert ^{-1} \rho_u(\xi),
	\label{eq:def_L_u}
\end{eqnarray}
where $f_u'(\xi)$ denotes the derivative of $f_u$ at $\xi$. Since the expanding rate of $f_u$ is the stability multiplier of $f$, the following holds:
\begin{eqnarray}
	\prod_{k = 0}^{t - 1} \left\vert f_u' (f_u^k(\xi)) \right\vert ^{-1} = e^{-\Lambda_1(\mathbf{x}, t) t},
	\label{eq:1d_2d_lyapunov}
\end{eqnarray}
where $\Lambda_1(\mathbf{x}, t)$ is the mFT Lyapunov exponent of $f$. Since the map $f_u$ defined on $(-\infty, \infty)$ is monotonically expanding, $\vert \mathcal{L}_u^{t} \rho _u(f^t(\xi)) \vert \rightarrow 0$ should follow as $t \to \infty$, consistently with the decay of $\rho(\mathbf{x})$ to the equilibrium distribution $\rho_\infty=1$. 

We hereafter consider the time evolution of the distribution function $\rho(\mathbf{x})$, which is assumed to be smooth and localized around an arbitrary point $\mathbf{x} = \mathbf{x}_0 \in \mathcal{M}$ and properly normalized in $\mathcal{M}$. Since the unstable manifold is dense in $\mathcal{M}$, for any close neighborhood of $\mathbf{x}_0$, there exists a point $\mathbf{x}_1$ on the unstable manifold. Then we introduce the function $\rho_{u}(\xi)$, which is defined by restricting the function $\rho(\mathbf{x})$ onto the unstable manifold on which $\mathbf{x}_1$ is located. Here we denote the coordinate of $\mathbf{x}_1$ along the unstable manifold by $\xi_1$. 

As mentioned above, any smooth function tends to the equilibrium distribution as $t \to \infty$, so for any $\mathbf{x}_0$  the difference $\vert \mathcal{L}^t \rho (f^t(\mathbf{x})) - 1 \vert$ should also tend to zero as $t \to \infty$. Therefore if the relaxation towards equilibrium can be described as
\begin{eqnarray}
	\vert \mathcal{L}^t \rho (f^t(\mathbf{x}_0)) - 1 \vert
	 \sim 
	 \vert \mathcal{L}^t \rho (f^t(\mathbf{x}_1)) - 1 \vert
	 \sim  \vert \mathcal{L}_u^{t} \rho_{u}(f_u^t(\xi_1)) \vert , 
	\label{eq:2dim_1dim_pf}
\end{eqnarray}
we obtain 
\begin{eqnarray}
	\vert c \sigma^t  \psi (f^t(\mathbf{x}_0))\vert \sim e^{-\Lambda_1(\mathbf{x}_1, t) t} 
	\rho_{u}(\xi_1).
\end{eqnarray}
The constants $c, \sigma$ do not depend on $t$, and $\rho_{u}(\xi_1)$ is a value of the initial distribution $\rho_{u}(\xi_1)$ at $\xi = \xi_1$. Hence the following equivalence relation holds for any pair of points $\mathbf{x}_0^{(1)}, \mathbf{x}_0^{(2)} \in \mathcal{M}$, 
\begin{eqnarray}
	\Lambda_1(\mathbf{x}_0^{(1)}, t) \leq \Lambda_1(\mathbf{x}_0^{(2)}, t) \Leftrightarrow \vert \psi(f^t(\mathbf{x}_0^{(1)})) \vert \geq \vert \psi(f^t(\mathbf{x}_0^{(2)})) \vert.
	\label{eq:lyapunov_second_eigenfunction}
\end{eqnarray}
This explains a good agreement between the second eigenfunction for the Perron-Frobenius operator and the spatial distribution of the mFT Lyapunov exponent, as shown in figure \ref{fig:lyapunov_nu1} [Recall that, in our notation from section \ref{sec:mFTLEdist}, $\mathbf{x}_t=f^t(\mathbf{x}_0)$].

We now provide further evidence in support of the previous argument, by examining other examples of perturbed cat maps. In the case of $\nu=1$, the second-largest eigenvalue is unique, so the above argument can be applied. On the other hand, for $\nu=2$, there appear multiple second-largest eigenvalues having the same absolute value, hereafter denoted respectively by $\sigma_{2, \mathrm{real}}$, $\sigma_{2, \mathrm{comp}}$ and $\sigma^*_{2, \mathrm{comp}}$, and shown in figure \ref{fig:double_pcm_eigenval}(a).

\begin{figure}[H]
	\subfigure[\hspace{4cm}]{
		\begin{minipage}{0.5\hsize}
			\begin{center}
				\includegraphics[width = \hsize, bb= 0 0 460.8 345.6]{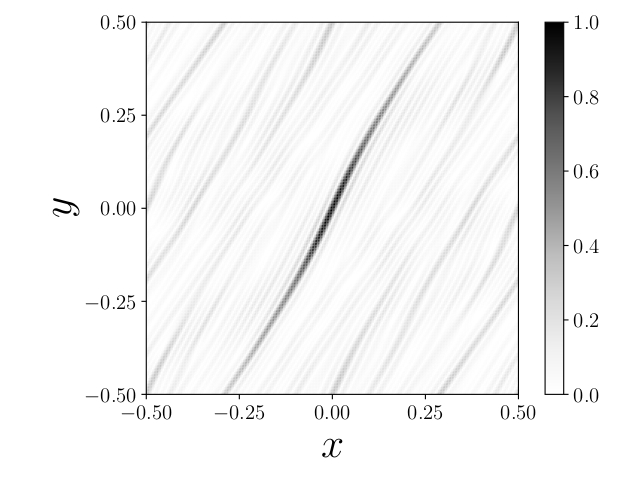}
			\end{center}
		\end{minipage}
	}
	\subfigure[\hspace{4cm}]{
		\begin{minipage}{0.5\hsize}
			\begin{center}
				\includegraphics[width = \hsize, bb= 0 0 460.8 345.6]{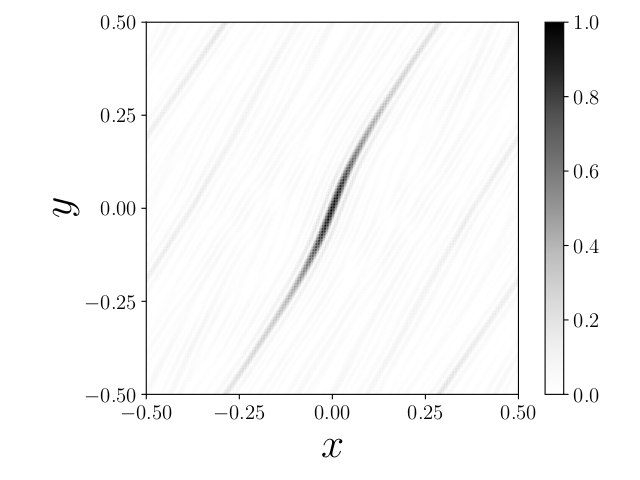}
			\end{center}
		\end{minipage}
	}
	\caption{Eigenfunction of the 
	Fokker-Planck operator for the map ${\cal C}_{\epsilon,\bar{\epsilon}}$ 
	associated with $\sigma_{2, \mathrm{real}}$
	(absolute value). (a)$\epsilon = 0.1$ and $\bar{\epsilon} = 0$. 
	(b)$\epsilon = 0.1$ and  $\bar{\epsilon} = 0.024$. We take $K=30$ and $\Delta = 10^{-3}$.}
	\label{fig:double_pcm_second_eigenfunc_real}
\end{figure}

\begin{figure}[H]
	\subfigure[\hspace{4cm}]{
		\begin{minipage}{0.5\hsize}
			\begin{center}
				\includegraphics[width = \hsize, bb= 0 0 460.8 345.6]{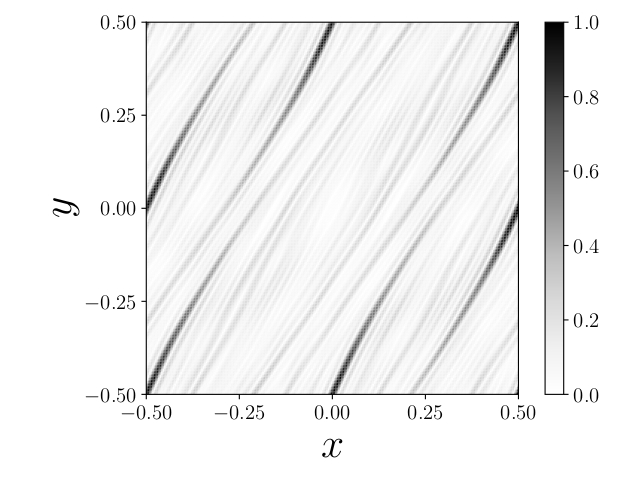}
			\end{center}
		\end{minipage}
	}
	\subfigure[\hspace{4cm}]{
		\begin{minipage}{0.5\hsize}
			\begin{center}
				\includegraphics[width = \hsize, bb= 0 0 460.8 345.6]{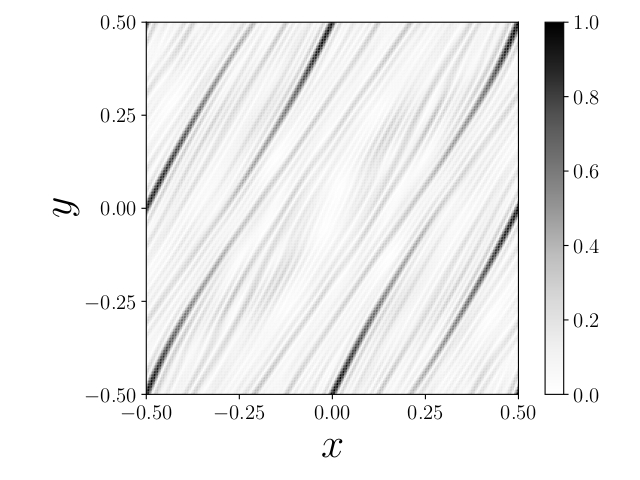}
			\end{center}
		\end{minipage}
	}
	\caption{
	Eigenfunction of the 
	Fokker-Planck operator for the map ${\cal C}_{\epsilon,\bar{\epsilon}}$ 
	associated with $\sigma_{2, \mathrm{comp}}$
	(absolute value). (a)$\epsilon = 0.1$ and $\bar{\epsilon} = 0$. 
	(b)$\epsilon = 0.1$ and  $\bar{\epsilon} = 0.024$. We take $K=30$ and $\Delta = 10^{-3}$. 
	}
	\label{fig:double_pcm_second_eigenfunc_comp}
\end{figure}

\begin{figure}[H]
	\subfigure[\hspace{4cm}]{
		\begin{minipage}{0.5\hsize}
			\begin{center}
				\includegraphics[width = \hsize, bb = 0 0 460.8 345.6]{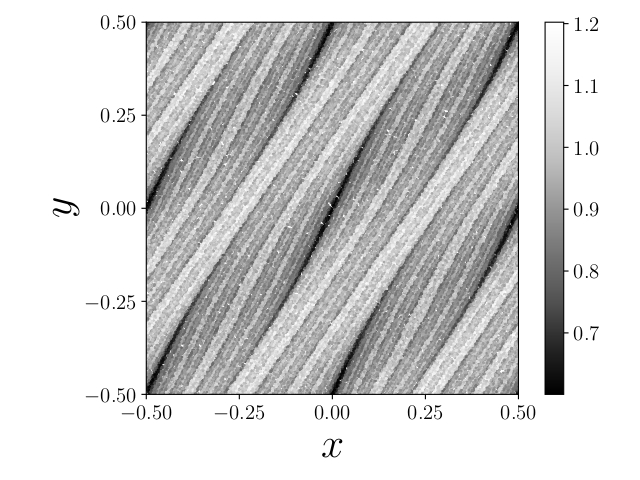}
			\end{center}
		\end{minipage}
	}
	\subfigure[\hspace{4cm}]{
		\begin{minipage}{0.5\hsize}
			\begin{center}
				\includegraphics[width = \hsize, bb = 0 0 460.8 345.6]{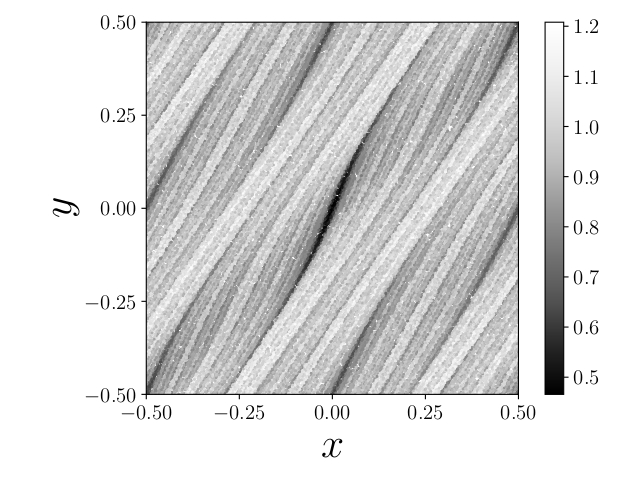}
			\end{center}
		\end{minipage}
	}
	\subfigure[\hspace{4cm}]{
		\begin{minipage}{0.5\hsize}
			\begin{center}
				\includegraphics[width = \hsize, bb = 0 0 460.8 345.6]{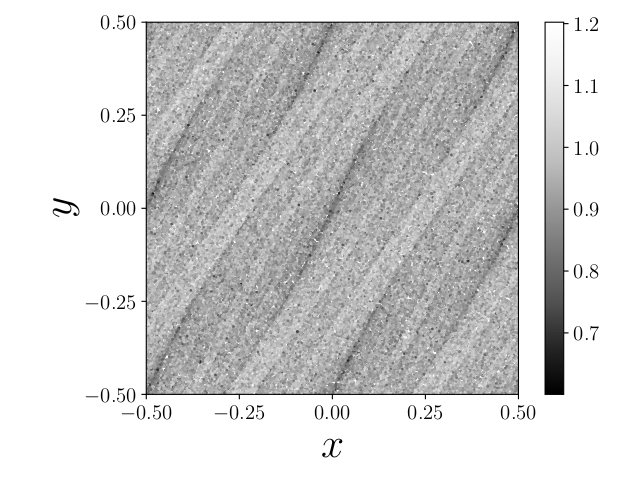}
			\end{center}
		\end{minipage}
	}
	\subfigure[\hspace{4cm}]{
		\begin{minipage}{0.5\hsize}
			\begin{center}
				\includegraphics[width = \hsize, bb = 0 0 460.8 345.6]{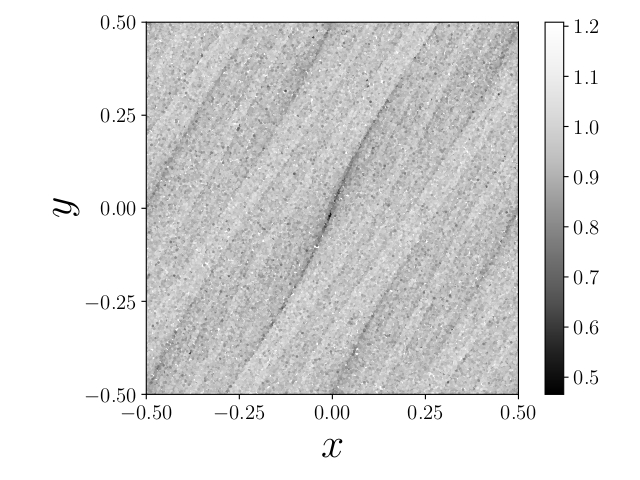}
			\end{center}
		\end{minipage}
	}
	\caption{Distribution of the mFT Lyapunov exponent of the map ${\cal C}_{\epsilon,\bar{\epsilon}}$.
	(a), (c) $\epsilon = 0.1$ and $\bar{\epsilon} = 0$. (b), (d) $\epsilon = 0.1$ and $\bar{\epsilon} = 0.024$. The time interval to calculate the mFT Lyapunov exponent is taken 
	as (a), (b) $t = 5$ and (c), (d) $t = 15$. The calculation is performed by sampling the phase space with $10^5$ randomly-distributed initial points.
		}
	\label{fig:double_pcm_lyapunov}
\end{figure}

For such a degeneracy, neither the localization pattern of the eigenfunction of the eigenvalue $\sigma_{2, \mathrm{real}}$ [figure \ref{fig:double_pcm_second_eigenfunc_real}(a)], nor that of the eigenvalue $\sigma_{2, \mathrm{comp}}$ [figure \ref{fig:double_pcm_second_eigenfunc_comp}(a)] single-handedly matches the spatial distribution pattern of the mFT Lyapunov exponent [figure \ref{fig:double_pcm_lyapunov}(a)]. However, when superposing the eigenfunctions of $\sigma_{2, \mathrm{real}}$ and $\sigma_{2, \mathrm{comp}}$, the resulting localization pattern agrees with that of the distribution of the mFT Lyapunov exponent. We notice that the latter might reflect a superposed state of two degenerated eigenfunctions, although degeneracy here occurs only in the sense of the modulus. We clarify this picture by introducing an additional perturbation to the perturbed cat map $C_{\epsilon,2}$, in order to lift the degeneracy of the second eigenvalues: 
\begin{eqnarray}
	{\cal C}_{\epsilon,\bar{\epsilon}} &= C_{\epsilon, 2} \circ {\cal F}_{\bar{\epsilon}}, \\
	{\cal F}_{\bar{\epsilon}}:(x, y) &\mapsto \left( x - \frac{\bar{\epsilon}}{2}\sin (4\pi y) \cos(2 \pi y), y \right). 
\end{eqnarray}
In the range $0 < \epsilon < 1/2\pi$, where the original perturbed cat map $C_{\epsilon, 2}$ is also hyperbolic, the larger eigenvalue of the Jacobian matrix of ${\cal C}_{\epsilon,\bar{\epsilon}}$ is real and greater than unity everywhere in the phase space, as long as the condition $0 < \bar{\epsilon} < 1/2\pi - \epsilon$ is satisfied. We will fix $\epsilon = 0.1$ in the following calculation.

We compute the Fokker-Planck operator of the map ${\cal C}_{\epsilon,\bar{\epsilon}}$. Unlike the original perturbed cat map $C_{\epsilon,2}$, the Fokker-Planck operator for the map ${\cal C}_{\epsilon,\bar{\epsilon}}$ does not have an analytic expression, so we need to carry out numerical integration in order to compute the matrix elements~(\ref{eq:fp_fourier}). We approximate the integral (\ref{eq:fp_fourier}) using the following summation:
\begin{eqnarray}
    \int_{\mathbb{T}^2}e^{2\pi \mathrm{i}(\mathbf{q}\cdot \mathbf{x}-\mathbf{k}\cdot C_{\bar{\epsilon}}(\mathbf{x}))}\mathrm{d}\mathbf{x} &\sim \sum_{i=0}^{M-1}\sum_{j=0}^{M-1}e^{2\pi \mathrm{i}(\mathbf{q}\cdot \mathbf{x}_{ij}-\mathbf{k}\cdot C_{\bar{\epsilon}}(\mathbf{x}_{ij}))} \frac{1}{M^2},
\end{eqnarray}
where $M$ denotes the lattice number in the discretization of the phase space, and $\mathbf{x}_{ij} = \left( \frac{i}{M}, \frac{j}{M} \right) (i,j = 0, \dots, M - 1)$ represents the $(i,j)$-th lattice point. 

As illustrated in figure~\ref{fig:double_pcm_eigenval}, the three-fold degeneracy in magnitude of the second-largest eigenvalues is lifted with the perturbation $\bar{\epsilon}$. Note that the real eigenvalue $\sigma_{2, \mathrm{real}}$ gradually increases while staying real-valued, whereas the magnitude of $\sigma_{2, \mathrm{comp}}, \sigma_{2,\mathrm{comp}}^{*}$  decreases, and the magnitude of the third-largest one $\sigma_{3}(\bar{\epsilon})$ exceeds that of $\sigma_{2, \mathrm{comp}}, \sigma_{2,\mathrm{comp}}^{*}$ around $\bar{\epsilon} \sim 0.024$, as seen in figure~\ref{fig:double_pcm_eigenval}(d). In addition, figure~\ref{fig:double_pcm_eigenval}(b) shows that the third-largest eigenvalue $\sigma_{3}(\bar{\epsilon})$ increases as a function of $\bar{\epsilon}$ as well, nevertheless $\sigma_{2, \mathrm{real}}$ and $\sigma_{3}(\bar{\epsilon})$ keep a finite distance. The result of this second perturbation $\bar{\epsilon} >0$ is therefore the appearance of a unique, isolated second-largest eigenvalue. 
When there exists a unique, isolated second-largest eigenvalue, we can expect the eigenfunction of the Perron-Frobenius operator to reflect the spatial distribution of the mFT Lyapunov exponent, as is confirmed in the case of $\nu=1$. 
Indeed, as shown in figure~\ref{fig:double_pcm_second_eigenfunc_real}(b), the eigenfunction of the Fokker-Planck operator associated with $\sigma_{2, \mathrm{real}}$ gives a similar spatial dependence of the mFT Lyapunov exponent distribution, which is presented in 
figure~\ref{fig:double_pcm_lyapunov}(b). Both the eigenfunction and the distribution of the mFT Lyapunov exponents are localized in the region along 
the unstable manifold emanating from the fixed point $(0,0)$, and, to a lesser extent, in the region around the period 3 orbit located at $(0.5,0.5)$ [the unstable manifold is split into six branches by the modulo operation, cf. figure~\ref{fig:double_pcm_second_eigenfunc_real}(b)]. Since $\vert \sigma_{2,\mathrm{comp}}\vert$ is close to $\vert \sigma_{2,{\mathrm{real}}} \vert$, 
the amplitude around the period 3 orbit is still higher than that in other regions.
On the other hand, as shown in figure~\ref{fig:double_pcm_second_eigenfunc_comp}, 
the eigenfunction associated with $\sigma_{2, \mathrm{comp}}$ 
shows a spatial pattern, which is different from that of the mFT Lyapunov exponent distribution. 
This result further supports our claim that 
the second eigenfunction of the Perron-Frobenius operator reflects 
the spatial distribution of the mFT Lyapunov exponent. 

In the perturbed situation analyzed above, the second eigenvalue is real-valued, 
and the corresponding eigenfunction localized around the fixed point, 
which has the minimal instability. 
However, we can realize the situation in which the complex-valued eigenfunctions 
become the second-eigenfunction after resolving the degeneracy by 
flipping the sign of the perturbation $\bar{\epsilon}$ from positive to negative. 
In this case, the second eigenfunction looks almost the same as the one 
presented  in figure~\ref{fig:second_eigenfunc_localize_period3}. 
This means that the second eigenfunction is not 
necessarily localized around the periodic orbit (fixed point) that has the smallest local instability, 
as already observed in section \ref{sec:inhomgeneity_unst_mani}.

%% file: 7_conclusion.tex
We have numerically investigated eigenfunctions associated with the leading sub-unit eigenvalues of the Perron-Frobenius operator in 2-dimensional uniformly hyperbolic area-preserving maps. We call them the second eigenfunctions and the second-largest eigenvalues, respectively. In the present paper, we have considered the cat map and the perturbed cat map as model systems, since these are known to be uniformly hyperbolic within some parameter range.

The eigenfunctions of the Perron-Frobenius operator for uniformly hyperbolic systems are expected to belong to a class of nontrivial hyperfunctions and functional spaces: one needs special care when calculating them numerically. The Ulam method is a well known and simple numerical technique to obtain eigenvalues and the corresponding eigenfunctions, and thus  it has been used as a numerical scheme in the physics literature \cite{ES2010, ES2012, KW2016}.

As demonstrated in section~\ref{sec:ulam}, however, numerically calculated second eigenfunctions for the unperturbed cat map provide inconsistent spatial patterns, depending on the choice of partitions constructing the transfer matrix. One might think it more reasonable to take Markov partitions in the computation of the transfer matrix of the Ulam method because the Markov partition is compatible with the dynamics, but it has been shown that the second-largest eigenvalue is sensitive to the type of Markov partition \cite{BSTV1997}. This implies that associated eigenfunctions also depend on the choice of the partition, and this is actually confirmed in section~\ref{sec:ulam}, although the spatial patterns seem to stabilize as the partition is made finer. The same computation is  generally unstable when the uniform partition is used to evaluate the transfer matrix. The second-largest eigenvalue does not converge, and spatial patterns of the second eigenfunction vary sensibly as a function of the grid resolution, if we adopt the uniform partition. From the observations presented here, we must conclude that the Ulam method is not reliable for the unperturbed cat map.

The Ulam method is shown to be relatively efficient and seems to work reasonably well when it is applied to the perturbed cat map. In particular, for a large enough nonlinear perturbation, we obtain better convergence for the second-largest eigenvalue. What is interesting is that the spatial distribution of the second eigenfunction is highly inhomogeneous, and localized in some regions of the phase space. Such localization reminds us of quantum scarring, which is found in quantum eigenfunctions of chaotic systems, thus it might be called classical scarring although it is not clear to what extent they share a common origin~\cite{LSYY2020}.

Next, we have investigated the second-largest eigenvalues and the associated eigenfunctions of the Perron-Frobenius operator by calculating the Fokker-Planck operator with sufficiently small diffusivity. As shown in section \ref{sec:fokkerplanck}, this approach works well and more stable compared to the Ulam method: as the maximum wavenumber included in constructing the Fokker-Planck kernel is increased, well convergent second-largest eigenvalues and second eigenfunctions are obtained for each value of diffusivity. Although the zero diffusivity limit is subtle and singular, we can infer from these observations that the second-largest eigenvalue take a finite value and the associated eigenfunction has inhomogeneous spatial signatures in case of the perturbed cat map. 

Spatial inhomogeneity or localization of the second eigenfunctions, found by applying the Ulam method, is also observed with the Fokker-Planck operator approach. We verify that second eigenfunctions are both localized around the region where the unstable manifold emanating from a fixed point is sparsely distributed. Orbits are less likely to enter sparse regions of unstable manifolds, and conversely it takes longer for trajectories to escape from such regions, compared to everywhere else in the phase space. Note that such a position dependence of the relaxation rate appears previous studies of the escape rate \cite{KL2009, BY2011, Bun2012, APT2013, BD2007, AB2017}.

Here for the perturbed cat map we discovered that the distribution of the mFT Lyapunov exponents exhibit quite a similar spatial pattern as that of the absolute value of the second eigenfunction. The mFT Lyapunov exponent measures the instability of the dynamics within a finite time interval, and thus depends on the initial (or final) position in phase space. We need to set an optimal time interval for which good correspondence is found between the distribution of the mFT Lyapunov exponents and the patterns of the second eigenfunction: for one-step iteration, the mFT Lyapunov exponent measures the local instability of the map. On the other hand, the asymptotic (infinite-time) calculation provides the conventional Lyapunov exponent, which is almost everywhere constant and does not depend on the initial position in the phase space, since the system is uniformly hyperbolic.

A close connection between spatial patterns of the second eigenfunction and the mFT Lyapunov exponent distribution might be feasible or intuitively acceptable because the second eigenfunction is expected to reflect the spatial dependence of the relaxation for a long time, whereas the mFT Lyapunov exponent controls the position-dependent instability for a certain time interval. However, we finally point out that there exists a counterexample where this hypothesis fails. Such a map can be obained by modifying the quadbaker map introduced by Froyland \cite{Fro2007}. In the modified quadbaker map, despite the fact that the local instability is completely uniform and the same is true for the mFT Lyapunov exponent, the second eigenfunction takes several different values in phase space, so not a constant function. We do not have a conclusive explanation for the mismatch between the spatial pattern of the second eigenfunction and that for the mFT Lyapunov exponent in this case, but it should be recalled that the cat map and perturbed cat map are both torus diffeomorphisms, while this modified quadbaker map is singular at boundaries of the phase space because the phase space is cut into pieces and then each piece is glued together under the dynamics. This might give rise to some other mechanism which makes the spatial pattern of the second eigenfunction nonuniform.